\documentclass[appendixfloats,iop]{emulateapj}
\begin{document}

\title{Superluminous Spiral Galaxies}

\author{Patrick M. Ogle$^1$, Lauranne Lanz$^1$, Cyril Nader$^{1,2}$,  George Helou$^1$}

\affil{$^1$IPAC, California Institute of Technology, 
       Mail Code 220-6, Pasadena, CA 91125}

\affil{$^2$University of California, Los Angeles}

\email{ogle@ipac.caltech.edu}

\shorttitle{Super Spirals}
\shortauthors{Ogle et al.}

\begin{abstract}
We report the discovery of spiral galaxies that are as optically luminous as elliptical brightest cluster galaxies, with $r$-band monochromatic 
luminosity $L_r=8-14L^*$ ($4.3-7.5\times 10^{44}$ erg s$^{-1}$). These super spiral
galaxies are also giant and massive, with  diameter $D=57-134$ kpc and stellar mass $M_\mathrm{stars}=0.3-3.4\times 10^{11} M_\odot$. We find 53 
super spirals out of a complete sample of 1616 SDSS galaxies with redshift $z<0.3$ and $L_r>8L^*$. The closest example is found at $z=0.089$.
We use existing photometry to estimate their stellar masses and star formation rates (SFRs). The SDSS and {\it WISE} colors are consistent with normal star-forming spirals 
on the blue sequence. However, the extreme masses and rapid SFRs of $5-65 M_\odot$ yr$^{-1}$ place super spirals in a sparsely populated region of parameter space,
above the star-forming main sequence of disk galaxies.  Super spirals occupy a diverse range of environments, from isolation to cluster centers.  We find four super spiral 
galaxy systems that are late-stage major mergers--a possible clue to their formation. We suggest that super spirals are a remnant population of unquenched, massive disk galaxies. 
They may eventually become massive lenticular galaxies after they are cut off from their gas supply and their disks fade.

\end{abstract}

\section{Introduction}

The most massive galaxies in the universe are thought to form from the largest density peaks in the primordial  matter 
distribution.  Galaxy mergers change the initial galaxy mass function, forming more massive galaxies by combining less massive ones. The result is the galaxy mass 
distribution we see in the local universe, empirically described by the \cite{s76} luminosity function, together with a morphology-dependent mass-to-light ratio. 
The luminosity function also depends upon the star formation history of galaxies, regulated by gas content, gas accretion, stellar feedback, and active galactic nucleus
(AGN) feedback. Galaxy mergers play an important role here too, since tidal torques in merging systems force gas into the galaxy centers, leading to starburst activity that 
grows the stellar bulge and AGN activity that grows the supermassive black hole \citep{tt72,bh91,hcy09}. 

Galaxies segregate into two major classes based on color and morphology \citep{sik01, lss08}. Blue, star-forming disks (late-type galaxies, LTGs)  lie in one region of color-space 
called the blue-sequence. Red-and-dead spheroids (early-type galaxies, ETGs) lie in a different region of color space called the red sequence.  LTGs demonstrate a correlation
between star formation rate (SFR) and stellar mass ($M_\mathrm{stars}$) called the star-forming main sequence \citep[SFMS:][]{bcw04,edl07,wfv11}.  The SFMS may be a consequence of an equilibrium between 
inflowing gas and star-formation driven outflows, with the specific star formation rate (SSFR) regulated by the halo mass growth rate \citep{lcp13}. Most ETGs on the other hand have much
 lower SFRs because they lack the cold gas needed to sustain star formation. 
 
 It appears that there is a limit to the mass of star-forming disk galaxies of roughly $3 \times 10^{10} M_\odot$, 
 with the most massive disk galaxies transitioning away from the main sequence as their SSFR declines. This decline appears to be a gradual process, 
 occurring over a period longer than 1 Gyr after the gas supply to the galaxy disk has been interrupted \citep{sus14}. Rapid quenching does not appear to occur for most galaxies that 
 remain disk galaxies, contrary to early attempts to explain the apparently bimodal distribution of galaxy colors.

A number of mechanisms have been suggested to explain why the gas supply is interrupted for the most massive disk galaxies.  Major galaxy mergers
may disrupt merging disk galaxies and transform them rapidly into elliptical galaxies \citep{bgb04}, though this does not explain the transformation of galaxies that remain
disks. The accretion of cold gas onto a galaxy may be stopped when the galaxy halo becomes massive enough that accretion shocks develop, interrupting the cold streams
of gas needed to replenish the disk \citep{db06}.  Increasing AGN feedback from a growing supermassive black hole may shock or eject gas from the galaxy disk, reducing 
its capacity to form stars \citep{hhc06,o14}. Ram-pressure stripping of the interstellar medium (ISM) by the intercluster medium (ICM) of a galaxy cluster can also remove cold gas \citep{srr14}.

Studying the most massive spiral galaxies can give us clues as to which of the above evolutionary processes are primarily responsible for converting star-forming
disk galaxies into red-and-dead lenticulars or ellipticals. The existence of rapidly star-forming, massive spirals with $M_\mathrm{stars}>10^{11} M_\odot$ 
indicates that disk galaxies can postpone this fate under special circumstances.  
We present here the most optically luminous and biggest spiral galaxies at redshift $z<0.3$, found by mining the NASA/IPAC Extragalactic Database (NED).
We assume a cosmology with $H_0=70$, $\Omega_m=0.3$, and $\Omega_\Lambda=0.7$ for computing
all linear sizes and luminosities.

\section{Sample}

This project is an offshoot of our work to determine the completeness of NED and explore its potential for systematic studies of galaxy populations
(Ogle, P. et al., in preparation). NED provides a unique fusion of multi-wavelength photometry from {\it Galaxy Evolution Explorer} ({\it GALEX}), Sloan Digital Sky Survey (SDSS), and 
the 2-Micron All-Sky Survey (2MASS), among others, which we augment by {\it Wide-field Infrared Survey Explorer}
({\it WISE}) photometry, that allows us to estimate stellar masses and SFRs. We compared the redshift distribution of galaxies in NED 
at $z<0.3$ to a model redshift distribution for the universe derived using a redshift-independent luminosity function, in order to estimate the spectroscopic 
completeness of NED. We used the \cite{s76} luminosity function fits of \cite{bhb03}, which are based on $\sim$150,000 SDSS galaxies 
as our benchmark. We adopt their characteristic absolute magnitude value of  $M^*-5 \log_{10} h =-20.44 \pm 0.01$ ($L^*=5.41\times 10^{43}$ erg s$^{-1}$ at 6200$\AA$) 
for the SDSS $r$-band luminosity function. The redshift limit was made large enough  to capture the rarest, most luminous galaxies, but not so large as to require consideration of 
redshift evolution in the luminosity function. 

\subsection{SDSS r-band Selection of the most Optically Luminous Galaxies}

SDSS is the largest source of spectroscopic redshifts, with a spectroscopic selection limit of $r=17.77$ (Strauss et al. 2002). We find that NED is complete over the SDSS 
footprint out to $z=0.3$ for galaxies with $L_r>8L^*$, the most optically luminous and massive galaxies in the low-redshift universe.  
Our sample is chosen from the 797,729 galaxies (type$=$G) in NED with spectroscopic redshifts $z<0.3$, in the SDSS footprint, and detected in SDSS $r$ band. We apply 
Galactic extinction corrections (tabulated by NED) and K-corrections to the $r$-band magnitudes prior to making our sample selection. We find 1616 galaxies with  
redshift $z<0.3$ and luminosity $L_r>8L^*$, which constitute our Ogle et al. Galaxy Catalog (OGC). The most luminous galaxy in  the OGC is a  $20L^*$ elliptical 
brightest cluster galaxy (BCG). 

\subsection{UV Selection Method for Super Spiral Galaxies}

We make a further selection for UV emission because we are interested in finding the most massive, actively star-forming disk galaxies. 
We recently matched and integrated the {\it GALEX} All-Sky Survey Catalog (GASC) and {\it GALEX} Medium Sky Survey Catalog (GMSC) with NED, using an automated, 
statistical algorithm \citep{o15}.  We inspected the SDSS images of all 196 galaxies from the OGC that are detected in the {\it GALEX} NUV band (the OGC-UV subsample). 
Of these, we find 46 NUV-detected,  $L_r>8L^*$ galaxies with spiral morphology (Table 1). The remaining NUV sources include 118 ellipticals, 
11 galaxies with E+A spectra, 2 quasi-stellar objects (QSOs) with extended emission, and 19 galaxies with erroneous redshifts or magnitudes.   The 
most luminous elliptical galaxy in OGC-UV is a $16L^*$ BCG,  while the most luminous spiral galaxy has $L_r=14L^*$.

\subsection{Morphological Selection Method for Super Spiral Galaxies}

We inspected the 310 brightest galaxies of the full OGC sample, those with  $L_r>10.5L^*$, to see if we are missing any spirals with UV selection. 
Of these, we classified 11 super spirals, 253 ellipticals, 38 galaxies with erroneous redshifts or magnitudes,  6 lenticulars, 1 irregular, and 1 E+A galaxy. This inventory 
includes 7 additional super spirals (Table 1) that are not in the OGC-UV sample. Of these, 4 have no {\it GALEX} sources nearby, and 3 others have nearby {\it GALEX} 
sources that should be matched in NED, but are not, possibly because of confusion. This shows that our NUV selection,  while relatively efficient (47/196) compared to morphological 
selection (11/310), leads to an incomplete sample, with only  4/11 super spirals recovered in this luminosity range. 
Part of the incompleteness (3/11) is owing to incomplete matching of {\it GALEX} with NED, while the rest (4/11) may be attributed to the {\it GALEX} detection limit or coverage.

\section{Photometry}

We conduct our investigation of super spirals primarily with photometry compiled by NED.
We use SDSS DR6 $u$, $g$, $r$, $i$, $z$ photometry measured with the CModel method, which combines exponential plus deVaucouleurs model fitting.
{\it GALEX} FUV and NUV photometry is taken from the GASC and GMSC, measured within a Kron elliptical aperture. 
We use 2MASS $J$, $H$, $Ks$ total magnitudes from the 2MASS Extended Source Catalog (2MASX).  NED objects are matched to AllWISE sources using the Gator tool 
in the Infrared Science Archive (IRSA). We use AllWISE 4.6 and 12 $\mu$m photometry within the largest available fixed-radius aperture of $24\farcs75$,
which is well-matched to the largest galaxy in our sample, with semimajor axis $a=24\farcs5$. 

\section{Basic Properties of Super Spirals}

\begin{figure*}
   \includegraphics[width=0.95\linewidth]{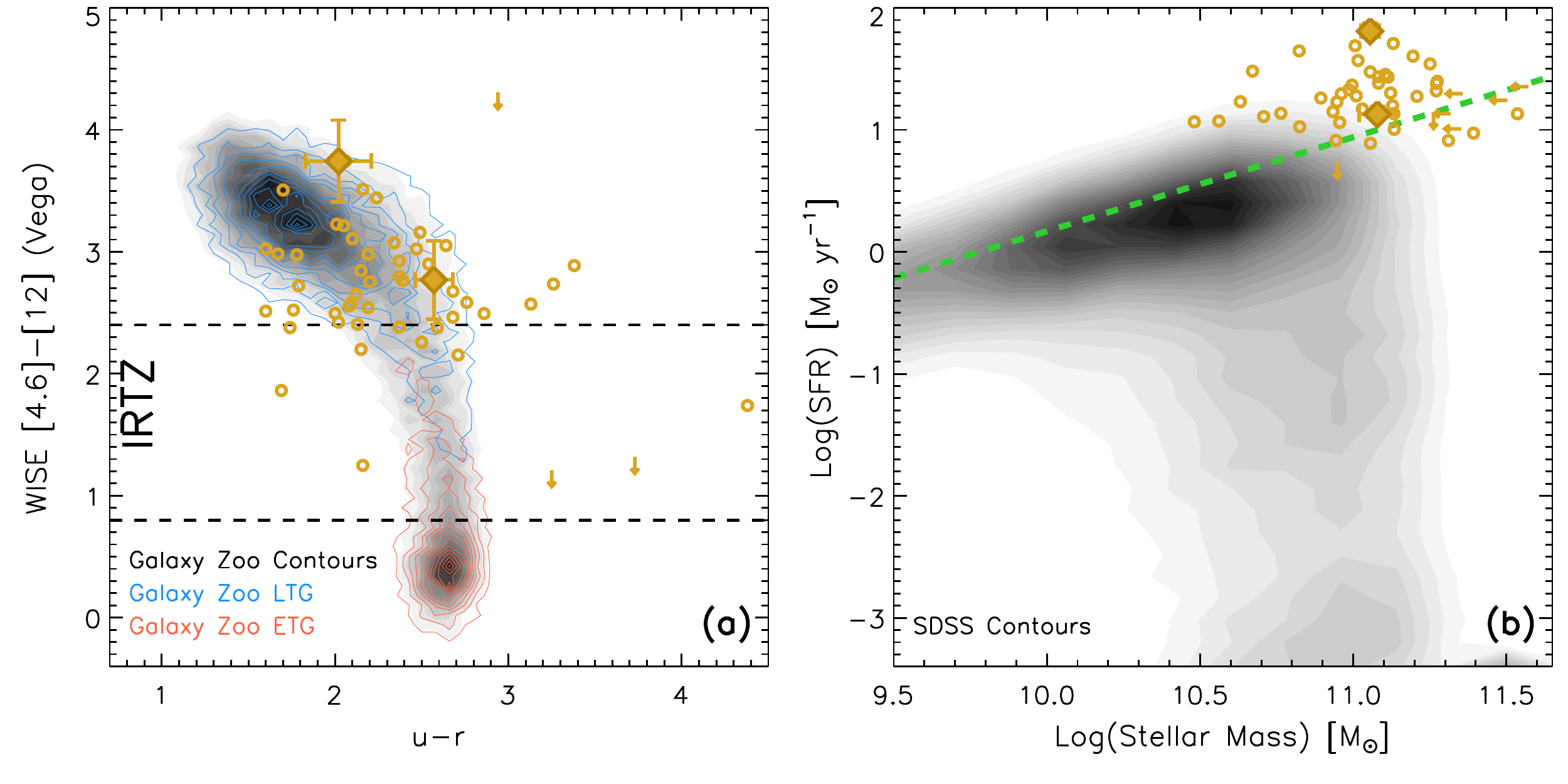}
   \figcaption{(a) SDSS and {\it WISE} colors of super spirals (circles) compared to other SDSS galaxies classified as LTG or ETG by \cite{lss08}. The infrared transition zone (IRTZ) is the mid-IR equivalent
   of the optical green valley \citep{asa14}. (b) Star formation rates and stellar masses of super spirals compared to the SDSS-{\it WISE} sample of \cite{cvc15}. The dashed line indicates 
   the star-forming main sequence at $z \sim 0$ \citep{edl07}. Galaxies above this line also have formation times that are generally less than the age of the universe. Larger diamond symbols are for  
   SDSS J094700.08+254045.7 (SS 16) and 2MASX J13275756+334529 (SS 05), with detailed SED analysis presented in the Appendix. 
   \label{1}}
\end{figure*}

\begin{figure}[t]
   \plotone{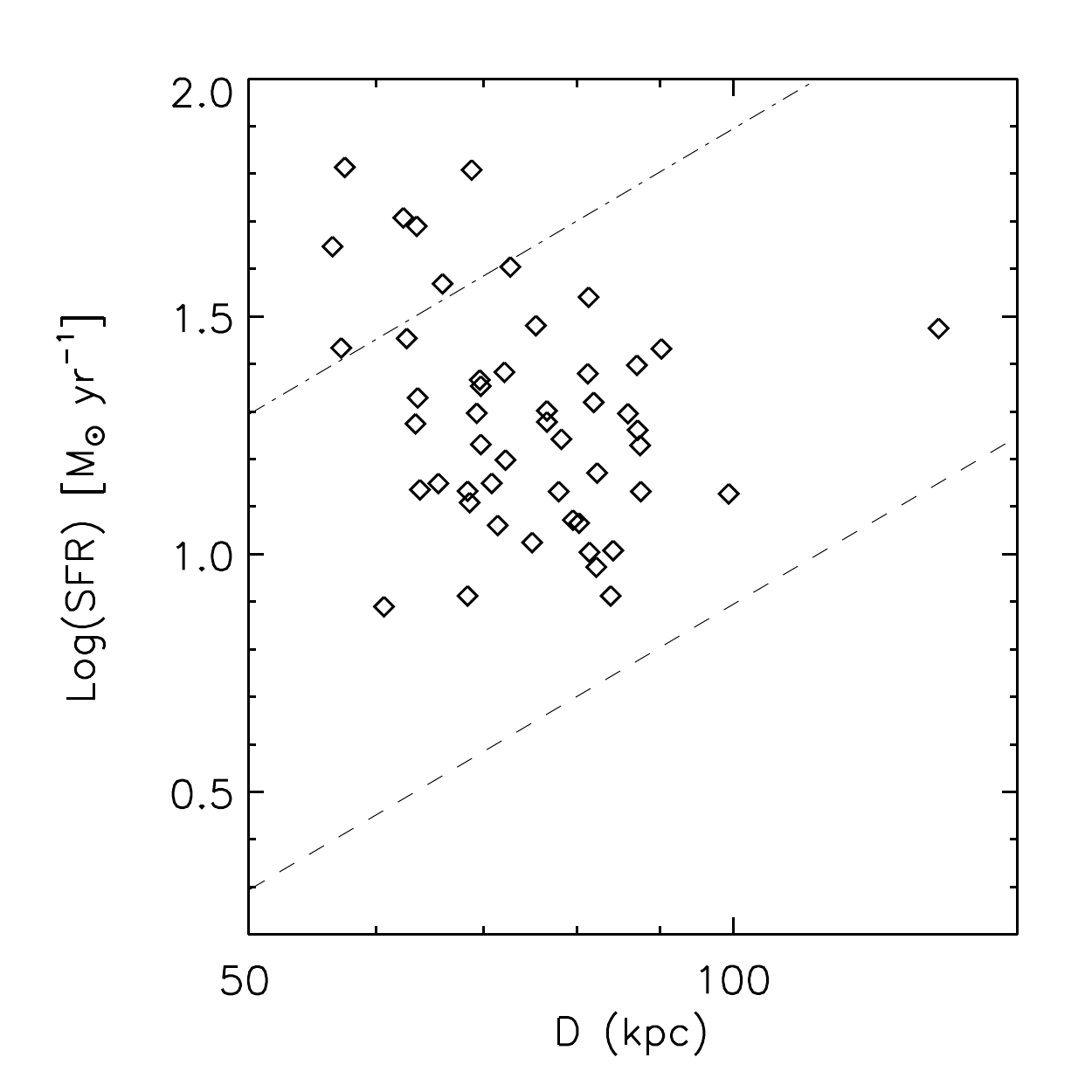}
   \figcaption{Super spirals range in diameter from 57 to 134 kpc.  The dashed and 
                     dotted-dashed lines indicate deprojected SFR surface densities of $1\times10^{-3}$ and $1\times10^{-2} M_\odot$ yr$^{-1}$ kpc$^{-2}$, respectively. 
                     \label{2}}
\end{figure}

\begin{figure}[t]
   \plotone{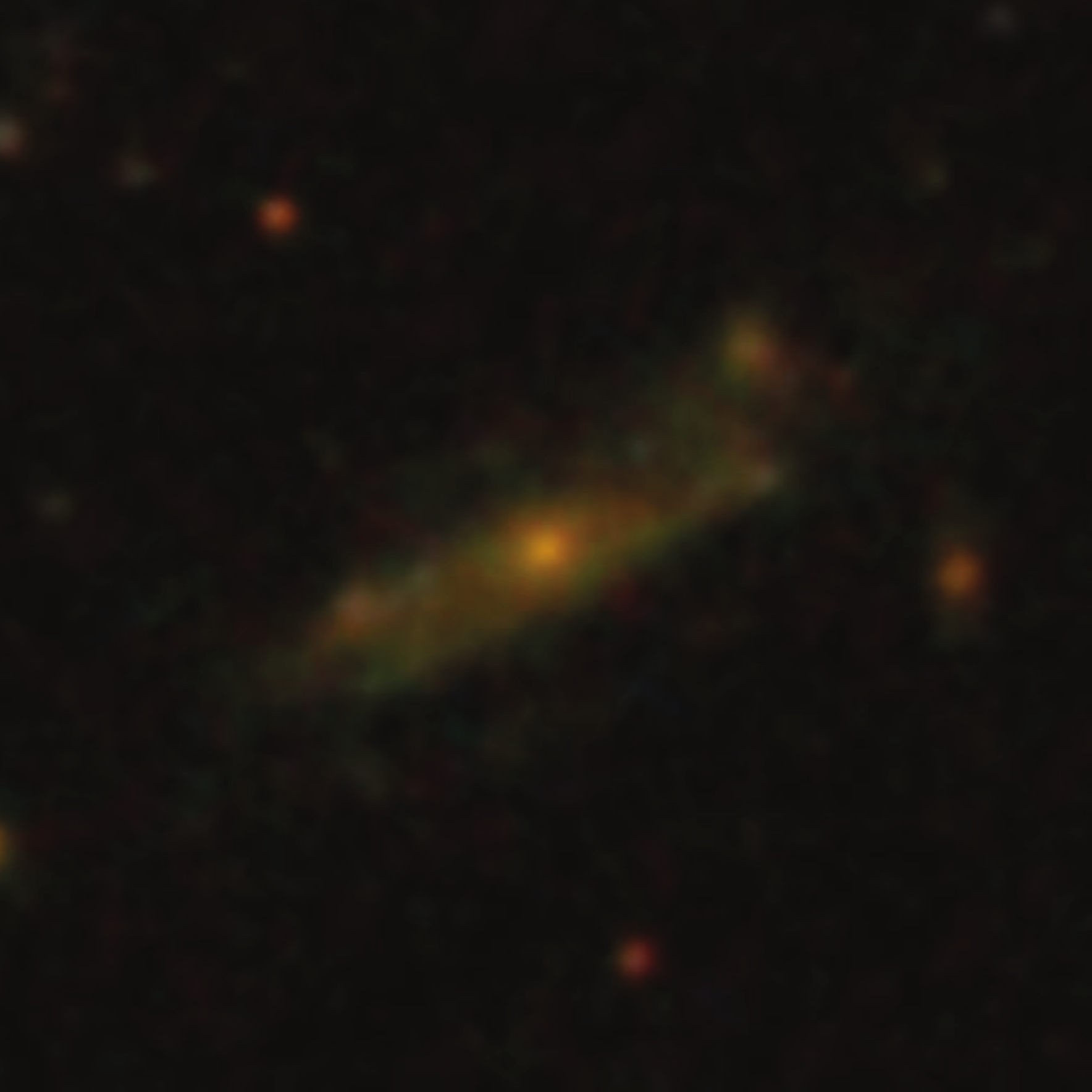}
   \figcaption{The largest super spiral galaxy, with $L_r=12L^*$ and an isophotal diameter of 134 kpc,  2MASX J16394598+4609058 (SS 03, $z=0.24713$). The SDSS image is 50.7$''$ (197 kpc) on each side. 
                     \label{3}}
\end{figure}

\subsection{Optical and Mid-IR colors}

The SDSS and {\it WISE} colors of super spirals lie along the blue sequence, similarly to less luminous star-forming disk galaxies (Figure \ref{1}(a)).
The SDSS comparison sample is adopted from \cite{asa14}, who show that LTGs and ETGs classified by Galaxy Zoo (GZ) \citep{lss08} are well-separated 
in {\it WISE} [4.6]$-$[12] vs. SDSS $u-r$ color space. The {\it WISE} [4.6]$-$[12] color ranges from 2.0 to 4.2, typical of polycyclic aromatic hydrocarbon (PAH) and warm 
dust emission from gas-rich, actively star-forming galaxies. The $u-r$ color ranges from 1.4 to 4.4, indicating star-forming disks with a range of SSFR or dust extinction. 
We estimate differential K-corrections of $\Delta(u-r)<0.2$ mag in the redshift range $z=0.1-0.3$, by convolving several  spectral energy distribution (SED) models  
(e.g., those in the Appendix) with the SDSS filter curves. These corrections are not large enough to explain the additional scatter in the observed $u-r$ colors of super spirals.  

There is a shift in the locus of super spiral colors compared to less-massive blue sequence galaxies. Super spirals tend to have redder $u-r$ and bluer [4.6]$-$[12] colors compared 
to the SDSS distribution. This could in principle indicate either lower SSFR or increased extinction. However, the high SSFR of our sample (Figure \ref{1}(b)) runs contrary to the first explanation. 
Six super spirals have $u-r>3.0$, a value not attained by less luminous SDSS LTGs.  The two reddest galaxies (SS 53 and SS 09) 
may be misclassified peculiar elliptical galaxies with prominent shells. CGCG 122-067 (SS 50) may be redder because of its double bulge. The other 4 are clearly spirals, and require
further investigation and custom photometry to determine the cause of their unusually red $u-r$ colors.

\subsection{Stellar Mass and Star Formation Rate}

\begin{figure*}
   \includegraphics[trim=0.0cm 2.5cm 0.0cm 0.0cm, clip, width=\linewidth]{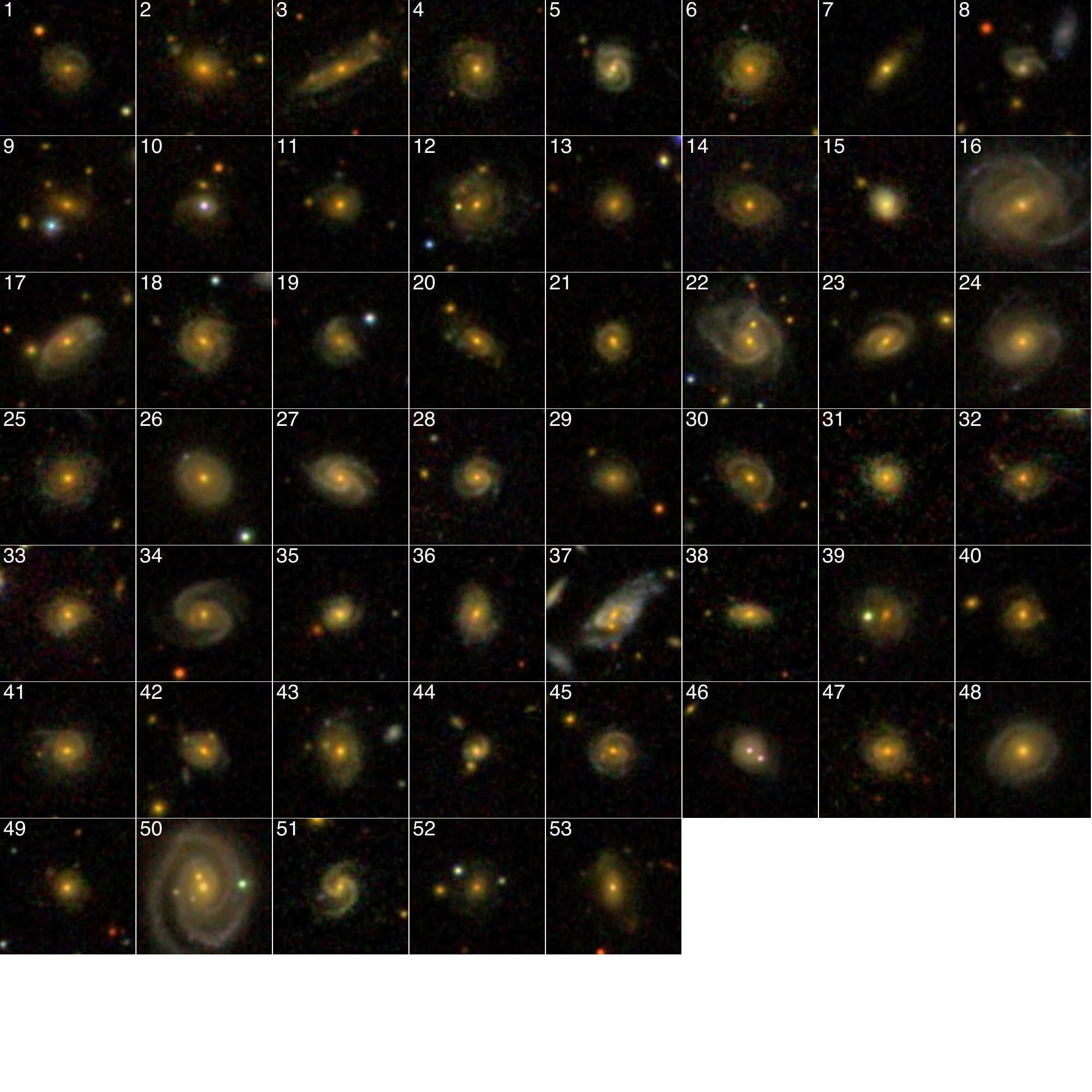}
   \figcaption{SDSS images of super spirals, 40$''$ on a side. Examples with peculiar morphology: (1) multi-arm spiral, (8) asymmetric 2-arm spiral,
    (10) QSO host with tidal arm, (21) ring galaxy, (23) possible tidal arm, (33) asymmetric disk, (34) possible secondary bulge, (53) partial arms or shells. 
    \label{4}}
\end{figure*}

We estimate stellar mass from 2MASS $Ks$ luminosity together with an SDSS $u-r$ color-dependent mass-to-light ratio 
estimated using the prescription of \cite{bmk03}, giving $M/L=0.75-1.34 M_\odot/L_\odot$. We apply a small correction to the stellar
masses to convert to a Chabrier initial mass function (IMF). This yields stellar masses that are consistent with more sophisticated SED template fitting (Appendix). 
We find stellar masses in the range $M_\mathrm{stars}=0.3-3.4\times 10^{11} M_\odot$. 

We estimate the SFR from the {\it WISE} 12 $\mu$m luminosity using the prescription of \cite{cvc15}, which was
established by SED-fitting more than 630,000 SDSS galaxies with {\sc magphys}  \citep{dce08}. While  
accurate for star-forming galaxies, this method may overestimate the star-formation rate for early type galaxies where dust
may be heated by other sources not directly related to star formation, or in the presence of a luminous AGN.  We further validate our
{\it WISE} single-band SFRs against  {\sc magphys} SED-fitting for two representative super spirals (Appendix). The {\it WISE} 12 $\mu$m monochromatic 
luminosities of super spirals range from $0.3-3.5\times 10^{44}$ erg s$^{-1}$ ($0.8-9.8\times10^{10}L_\odot$), corresponding to SFRs of $5-65 M_\odot$ yr$^{-1}$. 

We compare our sample to the SDSS-{\it WISE} sample of \cite{cvc15}, who estimated SFR and $M_\mathrm{stars}$ with {\sc magphys}. We find that most 
super spirals lie well above an extrapolation of the star-forming main sequence to higher mass (Figure \ref{1}(b)).  This is a region of the SFR 
vs. mass diagram that is very sparsely populated. The vast majority of SDSS disk galaxies in this mass range have significantly lower SFR and SSFR.

Our $r$-band luminosity plus NUV detection criteria tend to select galaxies with high global star formation rates. However, the SDSS spectra reveal
a relatively old bulge stellar population for most super spirals. We do find an indication of starburst activity in the SDSS bulge spectra of 3 super 
spirals (SS 05--see Appendix, SS 15, and SS 44) with strong young stellar population contributions and high-equivalent width H$\alpha$ emission. 
These three galaxies also have relatively blue SDSS $u-r$ colors and red {\it WISE} [4.6]$-$[12] colors, both indicative of a high global SSFR.

\begin{figure*}[t]
   \plotone{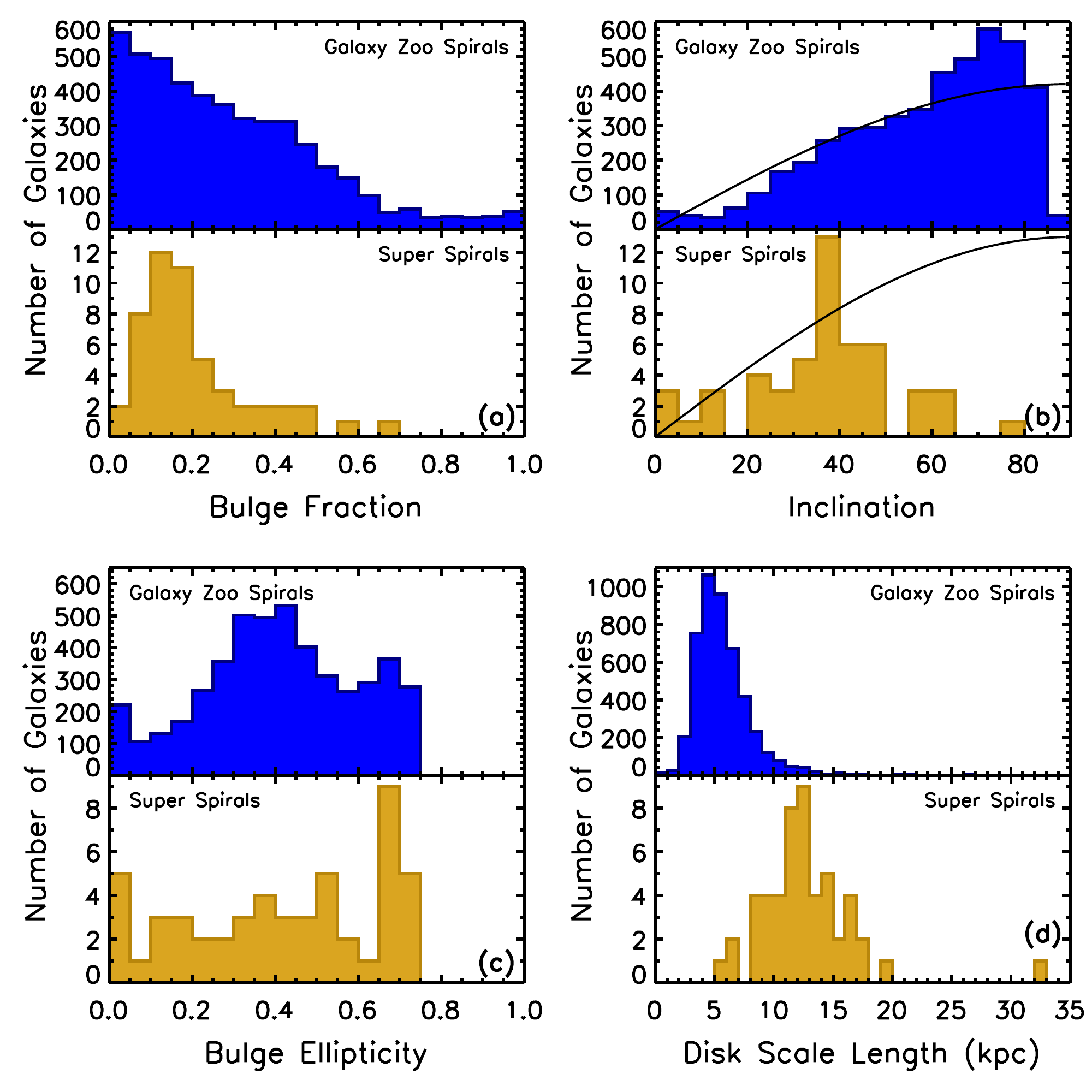}
   \figcaption{Distributions of super spiral and Galaxy Zoo spiral disk-bulge decomposition parameters, as measured by \cite{smp11}: (a) Bulge to total ($B/T$) $r$-band luminosity fraction,
                      (b) disk inclination distribution compared to the $\sin(i)$ expectation for randomly oriented disks (black curve), (c) bulge ellipticity, and (d) disk exponential scale length.
                      \label{5}}
\end{figure*}

\subsection{Active Galactic Nuclei}

The super spiral galaxies in our sample contain 3 Seyfert 1 nuclei and 2 QSOs with broad Balmer lines and strong [O {\sc iii}] in their SDSS spectra (Table 1). There is also 1 Seyfert 2 nucleus 
with strong [O {\sc iii}] but narrow Balmer lines. There is likely a dominant contribution from the QSO to the IR luminosity of  2MASX J15430777+1937522 (SS 10),
which has the greatest {\it WISE} 12 $\mu$m luminosity of our sample ($1.7\times10^{45}$ erg s$^{-1}$ or $4.3\times10^{11} L_\odot$).
The two QSOs are also detected at X-ray wavelengths by ROSAT.  One additional galaxy (2MASX J10095635+2611324 $=$ SS 45) is detected in X-rays, but
has no obvious signature of an AGN in its SDSS nuclear spectrum. There is so far no indication of any extended X-ray emission associated with super
spirals, though none have been specifically targeted for this. It will be important to make deep X-ray observations of super spirals to quantify any X-ray halo emission
in comparison to giant elliptical galaxies.  Only two super spirals are detected by the NVSS radio survey (2MASX J14472834+5908314 $=$ SS 47
and CGCG 122-067 $=$ SS 50), but the resolution is insufficient to distinguish between radio emission from star formation activity or from a radio jet.
The presence of luminous AGNs in 11\% of super spirals indicates that they are continuing to grow their supermassive black holes. 
It is imperative to measure the distribution of bulge and supermassive black hole masses in our super spiral sample
to see if they follow the same relation as lower-mass spiral bulges.

\subsection{Size, SFR Surface Density, and Morphology}

The sizes of super spirals  range from 57 to 134 kpc, with a median size of 72 kpc, using the SDSS DR6 $r$-band isophotal diameter at 25.0 mag arcsec$^{-2}$ (Table 1 and Figure \ref{2}).  
Their deprojected SFR surface densities range from  $1.5\times10^{-3}$ to $2.0\times10^{-2} M_\odot$ yr$^{-1}$ kpc$^{-2}$. A plot of SFR vs. diameter shows considerable 
scatter (Figure \ref{2}). However, the five most rapidly star-forming galaxies, with $\log(\mathrm{SFR}) > 1.6$, all have diameters $D<70$ kpc. The most MIR-luminous super spiral (SDSS J094700.08+254045.7, see Appendix), also has the largest deprojected SFR surface density. The largest super spiral, 2MASX J16394598+4609058 (SS 03, Figure \ref{3}), has a diameter of 
134 kpc  and a relatively low SFR surface density of $2.0\times10^{-3} M_\odot$ yr$^{-1}$ kpc$^{-2}$.

Super spirals display a range of morphologies, from flocculent to grand-design spiral patterns (Figure \ref{4}).  At least 9 super spirals have prominent stellar bars visible in the SDSS images (Table 1: Notes).  There are morphological peculiarities in several cases, including one-arm spirals, multi-arm spirals, rings, and asymmetric spiral structure (Figure \ref{4} and Table 1).  These
types of features may indicate past or ongoing galaxy mergers or collisions.

\subsection{Bulge-disk Decomposition}

We make use of the bulge-disk decompositions of \cite{smp11} to quantify the relative contributions of the bulge and disk to the luminosity of super spirals (Table 2). The
galaxy $g$ and $r$ band SDSS images are jointly fit by a de Vaucouleurs profile for the bulge (Sersic index $n_b=4$), plus an exponential disk. We compare super spirals to a 
representative subsample of 4686 spiral galaxies with $z>0.09$ classified by Galaxy Zoo (GZ, Figure \ref{5}), with bulge-disk decompositions also by \cite{smp11}.  We find a much narrower 
distribution of $r$-band bulge to total luminosity ($B/T$) for super spirals, with a median value of $B/T=0.17$, and a deficit of $B/T$ values $<0.1$.  A Kolmogorov-Smirnoff (K-S) test shows that the 
distributions differ significantly, with a  probability of only 0.0027 that super spirals are drawn from the same population as GZ spirals.  The lack of super spirals with 
$B/T<0.1$ may be consistent with a past history of significant merger activity.  The bulge ellipticity distribution of super spirals is not significantly different from that of GZ spirals (Figure \ref{5}(c)). 
We note that since the profile fits do not include a bar component, the ellipticities may be augmented by the presence of a bar or double bulge. 

The disk inclination distribution of super spirals also differs significantly from that of GZ spirals, and from the expected $\sin(i)$ dependence (Figure \ref{5}(b)).  A  K-S test gives a probability of
$\ll 0.001$ that super spirals and GZ spirals are drawn from the same inclination distribution. Only 5 (9\%) of super spirals have inclinations of $i>60\arcdeg$, compared to the expectation of 50\% for randomly oriented disks. This indicates that we are missing roughly 45\% of the super spirals in our luminosity range, possibly because of internal extinction at the NUV selection wavelength. The GZ spiral inclination distribution also differs from the expectation for randomly oriented disks, with an {\it excess} at inclinations $>60\arcdeg$ that may reveal a bias for GZ to classify edge on disks
as spirals or to misclassify edge-on lenticulars as spirals.

The median disk exponential scale length of super spirals is 12.2 kpc, 2.3 times as large as the 5.3 kpc median for GZ spirals, confirming the giant disk sizes of super spirals (Figure \ref{5}d).  
A  K-S test gives a probability of $\ll 0.001$ that super spirals and GZ spirals are drawn from the same size distribution. The galaxy smoothness parameter \citep{scd09}, which quantifies the
fractional residuals to the model fit inside two half-light radii,  is $S2=0.02-0.24$ in r band. The $B/T$ and $S2$ parameters of  bulge-disk decompositions have been used by others to 
quantitatively select early-type galaxies, with $B/T>0.35$ and $S2<0.075$ as criteria \citep{scd09}. Several super spirals in our sample meet these criteria, but we are nevertheless confident
of the detection of a significant spiral disk in most of these cases.

\section{Galaxy Merger Candidates}
\begin{figure*}
   \plotone{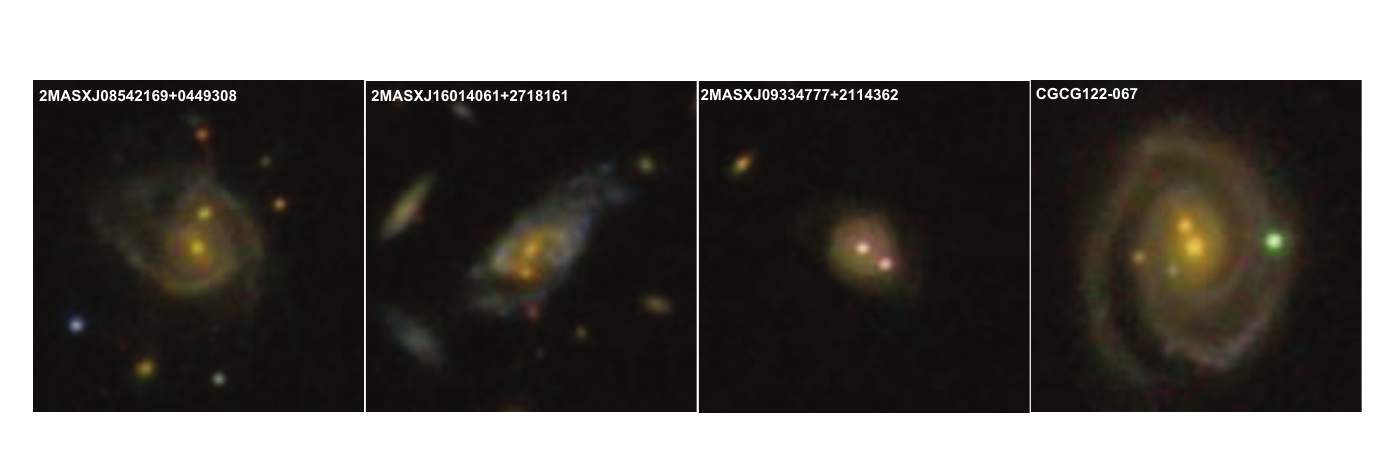}
   \figcaption{SDSS images of super spiral merger candidates. (a) Possible collision in progress of two spirals. (b) Possible 
   collision or merger of two spirals, also a brightest cluster galaxy (see also Figure \ref{7}). (c) High-surface
   brightness disk with possible double AGN, with faint outer arms. The nucleus at the center is classified as an 
    SDSS QSO. The second bright point source and possible AGN, near the edge of the disk, has a similar color to
    the primary AGN. (d) Possible late-stage major merger with two stellar bulges, with a striking grand spiral design
    surrounding both nuclei. Three other point sources may mark additional merging components or nuclei, reminiscent of nest
    galaxies commonly found at the centers of galaxy clusters. Each SDSS image is 48$''$ on a side.
     \label{6}} 
\end{figure*}

\begin{figure*}
   \plotone{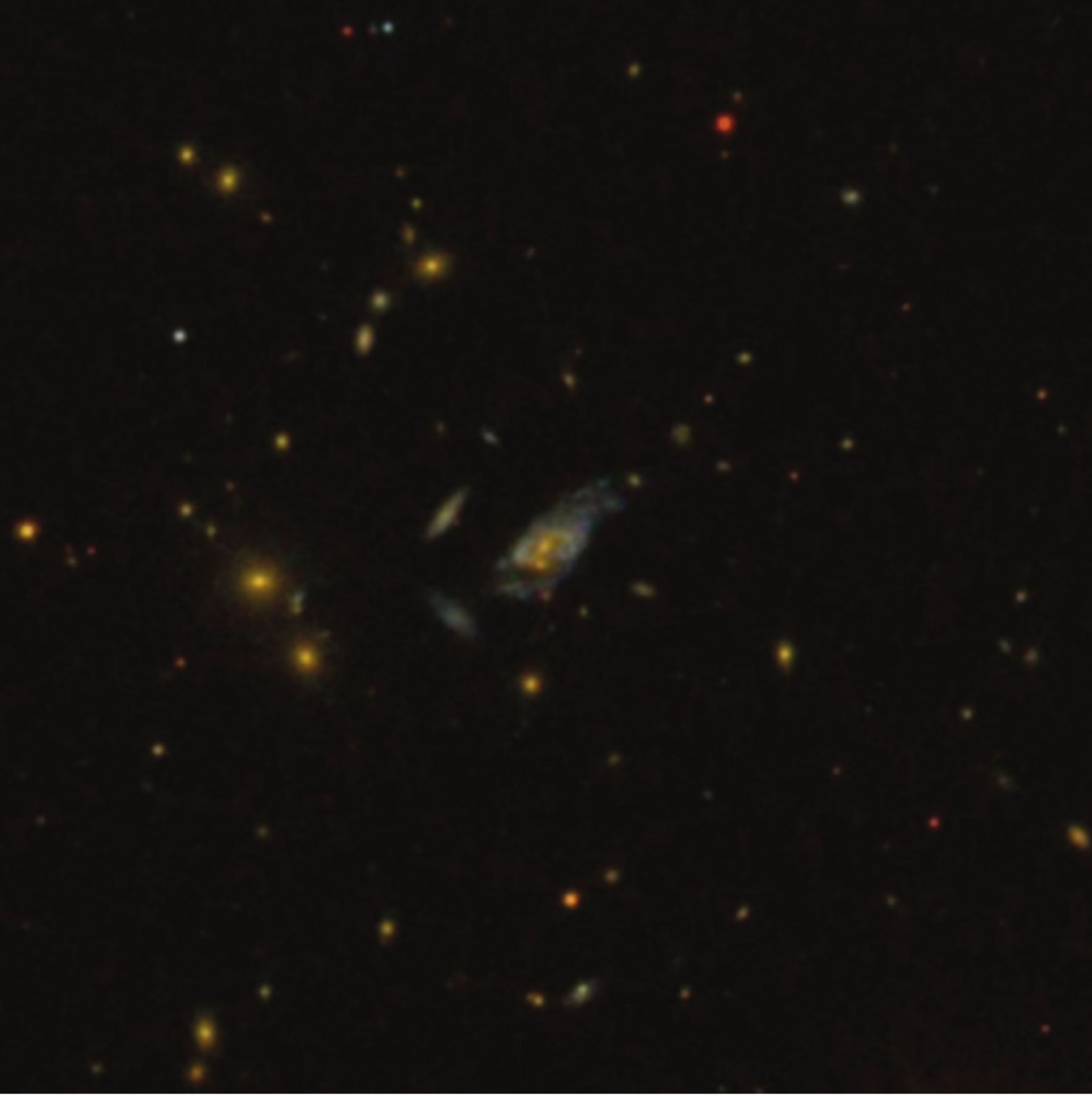}
   \figcaption{Super spiral merger candidate 2MASXJ16014061+2718161 (SS 37) is the brightest cluster galaxy of galaxy cluster GMBCG J240.41924+27.30444. The SDSS image
                     is $203''$ (572 kpc) on each side.
                     \label{7}}
\end{figure*}

\begin{figure*}
   \plotone{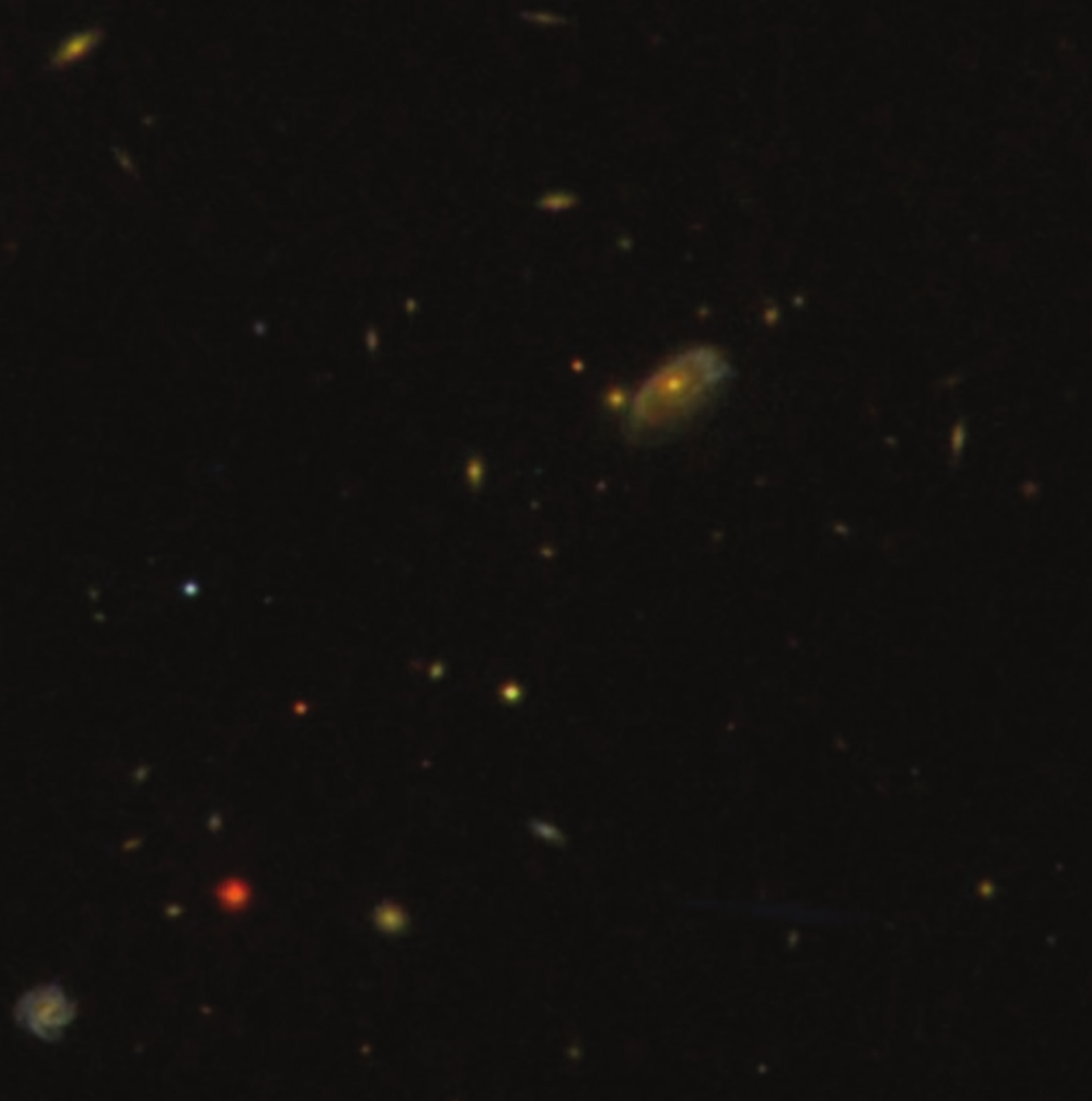}
   \figcaption{Super spiral 2MASX J11535621+4923562  (SS 17: $L_r=9.5L^*$, $D=90$ kpc) appears to be the brightest member of a previously unidentified galaxy cluster (OGC 0586 CLUSTER).  
                     Compare to the less-luminous cluster spiral galaxy SDSS J115407.96+492200.8 ($L_r=2.8L^*$, $D=39$ kpc) in the lower left corner. The SDSS image
                     is $203''$ (579 kpc) on each side.
                     \label{8}}
\end{figure*}

We find four super spiral merger candidates with apparent double stellar bulges or double nuclei (Figure \ref{6}).
The SDSS spectra only cover the dominant or central bulge or nucleus of each system. Spectroscopy of the secondary
bulges or nuclei will be necessary to confirm or rule out these merger candidates as true physical pairs or multiples.

The merger candidate 2MASX J08542169+0449308 (SS 22) appears to be a nearly equal mass major spiral pre-merger. The
arms of both spirals are wound in the same direction, and the disks appear to be overlapping in the plane of the sky.
The stretched out spiral arms of both spiral galaxy components, together with an apparent tidal arm at PA$=0\arcdeg$
(measured counterclockwise from North)
suggest an ongoing tidal interaction.

The merger candidate 2MASX J16014061+2718161 (SS 37) is a BCG, surrounded by several other
disk  galaxy companions (Figures \ref{6} and \ref{7}). 
The host cluster is identified as GMBCG J240.41924+27.30444, with a photometric redshift of 0.193 (Table 3). 
There are clear distortions to the spiral structures of both spiral galaxy components that appear to be involved in this merger.

The merger candidate 2MASX J09334777+2114362 (SS 46) appears to be a double AGN system. The primary, central nucleus
is identified as an SDSS QSO. The secondary nucleus has similar flux and color to the primary nucleus, but it does not have
an SDSS spectrum to confirm that it is a true physical double AGN. The galaxy disk has high surface brightness, suggestive
of starburst activity. Faint outer spiral arms are also suggestive of a recent galaxy interaction.

The merger candidate and BCG CGCG 122-067  (SS 50) appears to be a late stage $\sim2$:1 major merger.  The double bulge is surrounded
by a common inner disk. Two giant spiral arms emerge from this central disk, one from each bulge, making a complete
circuit around the disk. A  large gap is seen between the arms at PA$=0\arcdeg-90\arcdeg$. There are three other possible merging nuclei,
including a bright green point source at PA$=270\arcdeg$, that raise the possibility that this is a five-component multiple merger system.
Such multiple mergers are reminiscent of the elliptical nest galaxies that are sometimes found at the centers of galaxy clusters. 

\section{Environment}

We checked NED for known galaxy clusters and groups within $1'$ of each super spiral (Table 3). Seven of the super spirals are candidate BCGs, within $0.8''$ of 
a galaxy cluster. Two are candidate brightest group galaxies (BGGs), within $1'$ of a compact galaxy group. Most of the clusters only
have photometric redshifts and have yet to be verified spectroscopically. However, the photometric redshifts are all within $\Delta z = 0.04$ of the super spiral spectroscopic
redshift, which suggests a true physical association. The two associations of super spirals with compact groups are only based on their small angular separation,
with no independent redshift available for the groups.

We used NED's Environment Tool to further explore the environments of the super spiral BCG and BGG candidates. This tool performs a redshift-constrained cone search
for galaxies and galaxy clusters within a sphere of comoving radius 10 Mpc. Because of the high redshifts of the super spirals, only the most luminous galaxies
in their neighborhoods will tend to have measured spectroscopic redshifts in NED. We tabulate the number of galaxies (N1) with spectroscopic redshifts that are within 
1 Mpc and 500 km s$^{-1}$, and the number (N10) within 10 projected Mpc  and 5000 km s$^{-1}$. The MSPM 05544 galaxy cluster, which appears to host the super spiral
CGCG 122-067 (SS 50) has the largest number of cluster members with spectroscopic redshifts in NED (302), while the SDSSCGB 59704 galaxy group has the smallest number (2).
These numbers should be taken as lower limits to the cluster membership, depending primarily on the SDSS spectroscopic selection limit and redshift.

There are likely more clusters to be discovered in the vicinity of super spirals. For example, a clear overdensity of galaxies is seen to the SE of 
2MASX J11535621+4923562 (SS 17, Figure \ref{8}). We verify a concentration of 69 galaxies within 10 Mpc and  5000 km s$^{-1}$ (Table 3: OGC 0586 CLUSTER), using NED's Environment tool.
We estimate the mean redshift of OGC 0586 CLUSTER to be $z=0.166187$, from 12 galaxies with spectroscopic redshifts that are within 5  projected Mpc of SS 17.

\section{Discussion}

While super spirals have similar structure to less luminous spiral galaxies, they are impressive in the vastness of their scale. A sense of how truly enormous these galaxies
are can be gained by comparison to other galaxies in the same cluster (Figure \ref{8}: OGC 0586 CLUSTER). The 2MASX J11535621$+$4923562 (SS 17) super spiral at $z=0.16673$,
with a luminosity of $L_r=9.5L^*$ and a diameter of 90 kpc, can be compared to a more common, less luminous spiral galaxy which has $L_r=2.8L^*$ and a diameter of 39 kpc
($13\farcs7$), at about the same redshift ($z=0.16721$).

\subsection{Analogs}

It is natural to ask whether any analogs to super spirals have been found at lower redshift. One well-known example of a giant spiral galaxy is Malin 1 ($z=0.083$), initially suggested to be 
a proto-disk galaxy because of its massive H {\sc i} disk \citep{bim87}.  While Malin 1 does have one of the largest stellar disks known, with an exponential scale length of 70 kpc, its 
global $r$-band luminosity ($L_r=1.8L^*$) is not nearly great enough to make it into the OGC catalog. Its disk has very low surface brightness and is not readily visible in SDSS images. As 
further points of comparison, we estimate a global stellar mass of $1.2\times 10^{10} M_\odot$ and global SFR of $1.2 M_\odot$ yr$^{-1}$, which are both much lower than the range spanned 
by super spirals.

Other giant spiral galaxies are found in the local universe, though they also have considerably lower luminosities than the super spirals in our sample. \cite{r83} find 107 spiral galaxies
in the Uppsala General Catalog of Galaxies (UGC) at $z<0.05$, with cosmology-corrected $B$-band isophotal diameters (at $25.0$ mag arcsec$^{-2}$) of $D=65-150$ kpc, similar to super spirals. 
The 39 giant UGC spirals with SDSS photometry in NED have $r$-band luminosities of $0.2-4.6 L^*$, stellar masses of $6 \times10^8$ to $4 \times 10^{10} M_\odot$ yr$^{-1}$, and SFRs of 
$0.2-7.7M_\odot$ yr$^{-1}$.  Because of their considerably lower stellar masses, they cannot be faded super spirals, but could be useful analogs for understanding giant disks. One of the 
largest giant spiral galaxies, UGC 2885 has a rotational velocity of 280 km s$^{-1}$ at a radius of 60 kpc, and has undergone fewer than 10 rotations at its outer edge in the age of the universe
 \citep{rft80}.  

Super spirals may also be related to the cold sub-mm galaxies (SMGs) discovered at redshift $z=0.4-1$ \citep{csi02}. The relatively cold ($\sim 30 $ K) dust temperatures of 
these SMGs may indicate starburst activity in a disk rather than a spheroid. In comparison, the FIR SED of super spiral 2MASX J13275756+3345291 (SS 05, see Appendix)  is fit by the sum of 
a cold dust component with $T=21^{+0.9}_{-1.8}$ K, likely from the disk, and a warmer dust component with temperature $T=50^{+5.0}_{-2.8}$ K, likely from the starbursting bulge.
GN20, one of the most luminous sub-mm detected star bursting galaxies, shows molecular gas and star formation distributed in a 10 kpc scale disk at $z=4.05$ \citep{cdr10}. Deep near to 
mid-IR imaging of SMGs at intermediate redshifts will be necessary to measure their sizes and stellar masses and better determine their relationship to super spirals.

\subsection{Formation and Survival}

We estimate an average super spiral number density of $\sim 60$ Gpc$^{-3}$ at $z<0.3$, correcting for $\sim 45\%$ incompleteness at high disk inclination (\S 4.5). The
space density of super spirals is therefore only $\sim 6\%$ of the space density of elliptical galaxies in the same $r$-band luminosity range. Even the largest 
galaxy evolution simulations to date, such as the Illustris simulation \citep{vgs14,stl15}, covering $\sim1.0\times10^{-3}$  Gpc$^{3}$,  are not big enough to manufacture a significant number 
of super spirals. Therefore, no adequate prediction exists for the expected number of super spirals at $z<0.3$, nor are there simulations showing how these giant disk galaxies might form.

Super spirals could be formed  by gas-rich major spiral-spiral mergers.   Simulations that collide two gas-rich disk galaxies are able to produce post-merger spiral 
galaxies, albeit at smaller scale \citep{b02,sh05,rbc06,hcy09}. While merging stellar disks are typically destroyed, the gas in the outer disks may combine to reform an even larger gas and stellar
disk. Orbital geometry may also be important, with misaligned or retrograde orbits leading to more gas-rich final merger products. If the dynamical timescales are longer and the merger-induced 
torques are even weaker in the outer disks of super spiral mergers, this may also be conducive to the preservation of gas disks and reformation of stellar disks. 
Alternatively, super spirals might be formed more gradually, from the inside out by accretion of cold gas.  This may require a relatively low halo mass in order to avoid accretion shocks, which
might prevent the gas from settling onto the outer disk \cite{db06}. It will be important to study the spatial distribution of both neutral gas and star formation in super spirals to 
gain further insight into how their disks are formed.

It appears that the super spirals in our sample have so far avoided the fate of the vast majority of the most massive galaxies and continue to form stars in spite of their extreme mass, bucking the trend of
cosmic downsizing. There are several possible reasons for this success. First, super spirals may be robust to mergers because of their massive, dissipative gaseous disks. It appears 
that several super spirals in our sample have survived recent major mergers with their star-forming disks intact.  Second, the supermassive black holes in super spiral bulges may 
not be large enough to provide enough feedback to drive away the gas in the giant galaxy disk. Third, the halo mass may not be large enough to cut off cold accretion onto the disk
via accretion shocks. Finally, a large enough gas reservoir may have already settled into the disk to fuel star formation for a long time into the future.  Observations across the electromagnetic
spectrum are called for to distinguish among these possibilities.

\subsection{Connection to Quenched Disk Galaxies}

Super spirals occupy a relatively empty corner of the SFR vs. stellar mass diagram (Figure \ref{1}b). They lie above an extrapolation of the star-forming main sequence, at the most extreme mass 
and SFR. We find that most super spirals have SSFR $>0.08$ Gyr$^{-1}$. They are forming stars at a rate that would allow them to build up their mass in less than the age of the universe. 
This is unlike similarly massive, yet much more common disk galaxies (early type spirals and lenticulars) that fall below the star-forming main sequence, in what we shall call the disk quenching 
sequence (DQS: the disk galaxy subset of the green valley population). The disk-quenching sequence is discussed in the context of SSFR and UV color evolution by \cite{sus14}, and in the 
context of IR color evolution by \cite{asa14}.  Quenching disk galaxies are likely greatly reduced in their ability to form stars because their supply of cold gas has been cut off \citep[e.g.][]{db06}. 

The most densely populated ridge of the DQS is close to the median stellar mass of our super spiral sample ($M_\mathrm{stars}=1.1\times10^{11} M_\odot$).
We suggest that the majority of disk galaxies along this ridge were once super spirals. At a minimum, galaxies of this mass must have attained an average SFR$>7 M_\odot$ yr$^{-1}$ in order
to have formed in less than the age of the universe. This would put them squarely in the SFR and SSFR range of super spirals. A further implication is that their 
$D=60-130$ kpc diameter stellar disks must have faded dramatically. If fossil giant disks are detected around massive lenticular galaxies with deep imaging, it will provide strong
evidence for this hypothesis. In addition, deep H {\sc i} and CO observations may reveal if their cold gas reservoir has been entirely depleted or reduced to a level that is not conducive to 
star formation.

\section{Conclusions}

We report the discovery of a large sample of the most optically luminous ($L_r>8L^*$), biggest, and most massive spiral galaxies in the universe, which we call super spirals.
These galaxies are very rare ($\sim 60$ Gpc$^{-3}$ ), but are easily observed out to  $z=0.3$ because of their high luminosities and gigantic sizes.
Super spirals are forming stars at $5-65 M_\odot$ yr$^{-1}$,  a rate greater than their mean SFR over the age of the universe. 
Bulge-disk decompositions confirm the presence of giant stellar disks, with a median exponential scale length of 12.2 kpc, 2.3 times the median scale length
of less luminous spirals at the same redshift. The bulge-to-total optical luminosity distribution is also significantly different for super spirals, showing a deficit of galaxies
with $B/T<0.1$, and a concentration of galaxies with $B/T=0.1-0.2$.  Roughly 11\% of super spirals have Seyfert or QSO nuclei, suggesting that they are still 
actively adding mass to their supermassive black holes. We find evidence that several super spirals are undergoing major mergers, but manage to keep their 
star-forming disks intact, and avoid being transformed in to red-and-dead elliptical galaxies. Some super spirals are brightest cluster galaxies, while others appear to be 
isolated in the field. We suggest that super spirals may be the progenitors of red and dead lenticular galaxies of similar mass.

\acknowledgements
This work was made possible by the NASA/IPAC Extragalactic Database and the NASA/ IPAC Infrared Science Archive, which are both operated by the Jet Propulsion Laboratory, California Institute of Technology, under contract with the National Aeronautics and Space Administration.  We thank Joe Mazzarella, Ben Chan, Marion Schmitz, and the rest of the NED team for useful discussions and
their support of this work.  This publication makes use of data from the {\it Galaxy Evolution Explorer}, retrieved from the Mikulski Archive for Space Telescopes (MAST). STScI is operated by the Association of Universities for Research in Astronomy, Inc., under NASA contract NAS5-26555. Support for MAST for non-HST data is provided by the NASA Office of Space Science via grant NNX09AF08G and by other grants and contracts. Funding for the Sloan Digital Sky Survey IV has been provided by the Alfred P. Sloan Foundation, the U.S. Department of Energy Office of Science, and the Participating Institutions. SDSS-IV acknowledges support and resources from the Center for High-Performance Computing at the University of Utah. The SDSS web site is www.sdss.org.  This publication makes use of data products from the Two Micron All Sky Survey, which is a joint project of the University of Massachusetts and the Infrared Processing and Analysis Center/California Institute of Technology, funded by the National Aeronautics and Space Administration and the National Science Foundation. This publication makes use of data products from the {\it Wide-field Infrared Survey Explorer}, which is a joint project of the University of California, Los Angeles, and the Jet Propulsion Laboratory/California Institute of Technology, funded by the National Aeronautics and Space Administration.  This work is based in part on observations made with the {\it Spitzer Space Telescope}, which is operated by the Jet Propulsion Laboratory, California Institute of Technology under a contract with NASA. We also make use of data from {\it Herschel}, an ESA space observatory with science instruments provided by European-led Principal Investigator consortia and with important participation from NASA. We thank Katey Alatalo for providing the SDSS-{\it WISE} comparison data in Figure \ref{1}a, which is adapted from \cite{asa14}. We thank Phil Hopkins and Ski Antonucci for insightful discussions that contributed to the manuscript. 
Finally, we thank the anonymous referee for suggesting that we analyze available bulge-disk decompositions by \cite{scd09}, strengthening our results.

\eject

\clearpage

\begin{deluxetable}{cllccrrccccl}
\tablecaption{OGC Super Spiral Sample}
\tablehead{
\colhead{SS}  & \colhead{OGC} &\colhead{NED Name} & \colhead{$L_r$($L^*$)} & \colhead{$D$\tablenotemark{a}} &\colhead{$M_\mathrm{stars}$\tablenotemark{b}} &\colhead{SFR\tablenotemark{c}} & \colhead{Redshift\tablenotemark{d}} & \colhead{NUV}& \colhead{$r$} & \colhead{$u-r$} & \colhead{Notes}}
\startdata
01  & 0065 & 2MASX J10301576$-$0106068 & 13.9 & 81.3 &11.25&1.54&0.28228 & 21.02\tablenotemark{e}  & 16.92 &2.54& bar           \\
02  & 0073 & 2MASX J10405643$-$0103584  & 13.4 & 82.2 &11.39&0.97&0.25024 & 21.65 & 16.64 &2.16& BCG      \\
03  & 0139 & 2MASX J16394598$+$4609058  & 12.0 & 134. &11.05&1.48&0.24713 & 19.85 & 16.63 &2.37& edge-on\\
04  & 0170 & 2MASX J10100707$+$3253295  & 11.6 & 87.1 &11.27&1.40&0.28990 & 20.14 & 17.10 &2.68& BCG, bar   \\ 
05  & 0217 & 2MASX J13275756$+$3345291 & 11.2 & 68.8 &11.05&1.81&0.24892 & 19.44 & 16.72 &2.02& starburst, bar \\            
06  & 0256 & 2MASX J11593546$+$1257080  & 10.9 & 87.2 &10.89&1.26&0.26353 & 20.04 & 16.95 &1.79& \nodata \\    
07  & 0265 & SDSS J115052.98$+$460448.1 & 10.8 & 88.1 & 10.94&$<0.74$&0.28946 & $>$21.51\tablenotemark{e}& 17.19 &3.25& faint spiral \\
08  & 0290 & 2MASX J12343099$+$5156295  & 10.6 & 62.4 &11.13&1.71&0.29592 & 19.57 & 17.25 &1.67& Sy1, asymm.\\  
09  & 0299 & 2MASX J09094480$+$2226078  & 10.5 & 83.1 &11.26&$<1.15$&0.28539 & 21.40 & 17.25 &3.73& BCG, shells? \\
10  & 0302 & 2MASX J15430777+1937522 & 10.5 & 65.5 &11.37&2.45\tablenotemark{f}&0.22941 & \nodata & 17.07 & 0.40 &  QSO, tidal arm\\
11 & 0306 & SDSS J122100.48$+$482729.1 & 10.5 & 75.0 &10.82&1.02&0.29966 & 20.15 & 17.29 &1.69& \nodata \\   
12 & 0345 & 2MASX J09260805$+$2405242  & 10.3 & 81.2 &11.27&1.38&0.22239 & 19.61 & 16.57 &3.38& BCG, face-on  \\     
13 & 0388 & 2MASX J17340613$+$6029190  & 10.1 & 63.5 &11.20&1.27&0.27596 & 20.51 & 17.19 &2.71& BGG      \\   
14 & 0441 & SDSS J095727.02$+$083501.7 &  9.9 & 87.6 &11.53 & 1.13 &0.25652 & 20.88 & 16.99 &2.19& \nodata \\ 
15 & 0454 & 2MASXi J1003568+382901          & 9.9 & 56.4 &10.82&1.65&0.25860 & 19.79 & 16.97 &1.70& starburst       \\
16 & 0543 & 2MASX J09470010$+$2540462  &  9.6 & 99.3 &11.07&1.13&0.10904 & 17.74 & 14.83 &2.57& bar, Sy1?    \\    	 
17 & 0586 & 2MASX J11535621$+$4923562  &  9.5 & 90.2 &11.11&1.43&0.16673 & 19.92 & 15.90 &2.64& BCG, Sy2    \\ 
18 & 0595 & 2MASX J07550424$+$1353261  &  9.5 & 76.6 &11.12&1.30&0.22264 & 19.71 & 16.67 &2.47& bar      \\    
19 & 0696 & SDSS J102154.85$+$072415.5 &  9.2 & 69.7 &$<11.57$&1.35&0.29061 & 19.89 & 17.37 &2.01& \nodata \\ 
20 & 0713 & 2MASX J08265512$+$1811476  &  9.2 & 81.9 &11.27&1.32&0.26545 & 21.01 & 17.16 &3.26& bar \\   
21 & 0755 & SDSS J113800.88$+$521303.9 &  9.1 & 63.9 &10.76&1.14&0.29593 & 20.94 & 17.41 &2.12& ring   \\  
22 & 0789 & 2MASX J08542169$+$0449308  &  9.0 & 86.0 &10.96&1.30&0.15679 & 18.68 & 15.83 &2.49& 2 spirals, bar \\  
23 & 0799 & 2MASX J10472505$+$2309174  &  9.0 & 72.2 &11.12&1.20&0.18256 & 20.61 & 16.19 &2.68& bar      \\               	 
24 & 0800 & 2MASX J11191739$+$1419465  &  9.0 & 70.8 &10.93&1.15&0.14377 & 18.75 & 15.57 &2.39& \nodata \\  
25 & 0804 & SDSS J135546.07$+$025455.8 &  9.0 & 84.2 &$<11.35$&1.01&0.23884 & 19.77 & 16.87 &1.74& \nodata \\ 
26 & 0830 & SDSS J141754.96$+$270434.4 &  9.0 & 68.6 &10.70&1.11&0.15753 & 19.74 & 15.79 &2.86& \nodata \\ 
27 & 0926 & 2MASX J10304263$+$0418219  &  8.8 & 72.7 &11.19&1.60&0.16902 & 19.08 & 15.93 &2.16& \nodata \\     
28 & 0928 & 2MASX J12374668$+$4812273  &  8.8 & 66.0 &11.01&1.57&0.27245 & 19.79 & 17.24 &2.10& \nodata \\     
29 & 0975 & 2MASX J11410001$+$3848078  &  8.7 & 72.1 &11.08&1.38&0.26770 & 20.79 & 17.21 &2.15& \nodata \\    
30 & 0983 & SDSS J153618.97$+$452246.8 &  8.7 & 80.2 &10.48&1.07&0.23618 & 20.15  & 16.85 &2.13& \nodata \\
31 & 1046 & 2MASX J09362208$+$3906291  &  8.6 & 69.6 &10.99&1.37&0.28293 & 20.09 & 17.36 &1.78& \nodata \\ 
32 & 1088 & SDSS J140138.37$+$263527.6 &  8.5 & 78.2 &$<11.50$&1.24&0.28396 & 19.99 & 17.38 &2.08& \nodata \\
33 & 1196 & SDSS J154950.91$+$234444.1 &  8.4 & 69.3 &$<11.35$&1.30&0.26208 & 20.48 & 17.27 &2.02& \nodata \\ 
34 & 1250 & 2MASX J12321515$+$1021195  &  8.3 & 71.4 &10.95&1.06&0.16588 & 19.69 & 16.04 &2.76& 2 bulges?  \\     
35 & 1268 & 2MASX J12005393$+$4800076  &  8.3 & 62.7 &11.10&1.45&0.27841 & 20.04 & 17.37 &2.05& BCG      \\  
36 & 1273 & 2MASX J07380615$+$2823592  &  8.3 & 76.6 &11.01&1.28&0.23091 & 20.13 & 16.92 &2.34& \nodata \\  
37 & 1304 & 2MASX J16014061$+$2718161  &  8.3 & 82.3 &11.03&1.17&0.16440 & 17.60 & 16.06 &1.60& BCG, 2 spirals\\
38 & 1312 & SDSS J143447.86$+$020228.6 &  8.2 & 75.4 &10.67&1.48&0.27991 & 20.43 & 17.42 &2.24& \nodata \\ 
39 & 1323 & SDSS J112928.74$+$025549.9 &  8.2 & 69.7 &10.63&1.23&0.23960 & 19.56 & 17.01 &2.00& \nodata \\  
40 & 1352 & SDSS J101603.97$+$303747.9 &  8.2 & 68.8 &10.73& \nodata\tablenotemark{g} &0.25191 & 21.16  & 17.13 &2.94& \nodata \\
41 & 1375 & 2MASX J00155012$-$1002427 &  8.2 & 68.4 &10.94&0.91&0.17601 & \nodata              & 16.23 &2.09& \nodata \\
42 & 1395 & 2MASX J13103930$+$2235023  &  8.1 & 65.6 &11.08&1.15&0.23123 & 19.91 & 16.87 &2.59& \nodata \\	 
43 & 1420 & 2MASX J13475962$+$3227100  &  8.1 & 87.5 &10.94&1.23&0.22306 & 20.25 & 16.79 &2.37& BGG       \\       
44 & 1464 & 2MASX J10041606$+$2958441  &  8.1 & 57.4 &11.04&1.81&0.29844 & 20.64 & 17.59 &2.06& starburst \\ 
45 & 1500 & 2MASX J10095635$+$2611324  &  8.1 & 63.7 &10.98&1.33&0.24089 & 19.99 & 17.03 &2.19& \nodata \\  
46 & 1501 & 2MASX J09334777$+$2114362  &  8.1 & 63.6 &11.00&1.69&0.17219 & 17.84 & 16.17 &1.60& QSO, 2 nuclei\\
47 & 1544 & 2MASX J14472834$+$5908314  &  8.0 & 68.4 &11.13&1.13&0.24551 & 20.22 & 17.04 &2.15& \nodata \\     
48 & 1546 & 2MASX J13435549$+$2440484  &  8.0 & 60.7 &11.05&0.89&0.13725 & 19.54 & 15.56 &2.50& \nodata \\ 
49 & 1554 & 2MASX J13422833$+$1157345  &  8.0 & 57.1 &11.08&1.43&0.27873 & 21.66 & 17.43 &2.20& \nodata \\ 
50 & 1559 & CGCG 122-067                            &  8.0 & 81.4 &11.13&1.00&0.08902 & 18.27 & 14.56 &3.13& BCG, 2 bulges \\
51 & 1606 & SDSS J121644.34$+$122450.5 &  8.0 & 77.9 &$<11.31$&1.13&0.25694 & 20.12 & 17.22 &1.76& bar, Sy1   \\   
52 & 1608 & SDSS J040422.91$-$054134.9 &  8.0 & 79.5 &10.56&1.07&0.25055 & 20.38 & 17.27 &2.37& flocculent  \\
53 & 1611 & 2MASX J00380781$-$0109365  &  8.0 & 83.9 &11.31&0.91&0.20828 & 21.33 & 16.65 &4.38& E with shells?\\

\enddata
\tablenotetext{a}{Isophotal diameter (kpc) at $r=25.0$ mag arcsec$^{-2}$.}
\tablenotetext{b}{$\log_{10}M_\mathrm{stars}$ ($M_\odot$) or $3\sigma$ upper limit, based on 2MASS $Ks$ luminosity and $u-r$ color.}
\tablenotetext{c}{$\log_{10}$ SFR ($M_\odot$ yr$^{-1}$) or 95\% confidence upper limit, based on {\it WISE} 12 $\mu$m luminosity.}
\tablenotetext{d}{SDSS DR9 redshift.}
\tablenotetext{e}{{\it GALEX} NUV-band photometry measured in $14''$ aperture. Source not in GASC or GMSC.}
\tablenotetext{f}{This SFR may be overestimated by a large factor because of the QSO nucleus.}
\tablenotetext{g}{{\it WISE} data compromised by nearby IR-bright star.}
\end{deluxetable}

\clearpage

\begin{deluxetable}{clcccrcrc}
\tablecaption{Bulge-disk Decomposition \citep{smp11}}
\tablehead{\colhead{SS} & \colhead{NED Name}  & \colhead{Scale\tablenotemark{a}} & \colhead{$B/T$\tablenotemark{b}} 
&  \colhead{$e$\tablenotemark{c}} & \colhead{$R_d$\tablenotemark{d}}    & \colhead{$i$\tablenotemark{e}}  
&\colhead{PA\tablenotemark{f}} & \colhead{$S2$\tablenotemark{g}}  
}

\startdata
01 & 2MASX J10301576$-$0106068  & 4.268 & 0.13 & 0.67 & 14.39 & 37 &  33 & 0.07 \\
02 & 2MASX J10405643$-$0103584  & 3.914 & 0.66 & 0.25 & 12.43 &  2 &  74 & 0.04 \\
03 & 2MASX J16394598$+$4609058  & 3.879 & 0.13 & 0.01 & 32.98 & 76 &  28 & 0.04 \\
04 & 2MASX J10100707$+$3253295  & 4.349 & 0.12 & 0.22 & 15.31 & 31 &  57 & 0.02 \\
05 & 2MASX J13275756$+$3345291  & 3.898 & 0.29 & 0.70 & 11.34 & 40 &  77 & 0.17 \\
06 & 2MASX J11593546$+$1257080  & 4.063 & 0.03 & 0.03 & 16.24 & 35 & 114 & 0.04 \\
07 & SDSS J115052.98$+$460448.1  & 4.345 & 0.46 & 0.64 & 12.17 & 62 & 125 & 0.04 \\
08 & 2MASX J12343099$+$5156295  & 4.412 & 0.26 & 0.56 & 11.52 & 27 & -17 & 0.24 \\
09 & 2MASX J09094480$+$2226078  & 4.302 & 0.41 & 0.48 & 19.04 & 46 & 153 & 0.06 \\
10 & 2MASX J15430777$+$1937522 &\nodata & \nodata & \nodata & \nodata & \nodata & \nodata &  \nodata  \\
11 & SDSS J122100.48$+$482729.1  & 4.452 & 0.19 & 0.01 & 12.14 & 27 &  13 & 0.03 \\
12 & 2MASX J09260805$+$2405242  & 3.583 & 0.15 & 0.53 & 17.45 & 29 &  96 & 0.02 \\
13 & 2MASX J17340613$+$6029190  & 4.200 & 0.29 & 0.05 &  9.06 & 20 &  13 & 0.03 \\
14 & SDSS J095727.02$+$083501.7  & 3.986 & 0.13 & 0.42 & 16.97 & 39 &   4 & 0.06 \\
15 & 2MASXi J1003568$+$382901     & 4.009 & 0.11 & 0.24 &  6.25 & 39 &  50 & 0.06 \\
16 & 2MASX J09470010$+$2540462  & 1.991 & 0.21 & 0.51 & 17.31 & 43 & 150 & 0.08 \\
17 & 2MASX J11535621$+$4923562  & 2.855 & 0.17 & 0.38 & 15.67 & 63 & 144 & 0.13 \\
18 & 2MASX J07550424$+$1353261  & 3.585 & 0.07 & 0.69 & 12.91 & 24 &  87 & 0.07 \\
19 & SDSS J102154.85$+$072415.5  & 4.357 & 0.16 & 0.69 & 13.24 & 42 &  54 & 0.11 \\
20 & 2MASX J08265512$+$1811476  & 4.086 & 0.21 & 0.32 & 13.45 & 57 &   5 & 0.09 \\
21 & SDSS J113800.88$+$521303.9  & 4.414 & 0.23 & 0.50 & 12.83 & 36 & 103 & 0.11 \\
22 & 2MASX J08542169$+$0449308  & 2.713 & 0.08 & 0.68 & 12.14 & 36 &  58 & 0.10 \\
23 & 2MASX J10472505$+$2309174  & 3.071 & 0.18 & 0.70 &  8.84 & 45 & 156 & 0.17 \\
24 & 2MASX J11191739$+$1419465  & 2.523 & 0.46 & 0.11 & 10.22 & 47 & 153 & 0.07 \\
25 & SDSS J135546.07$+$025455.8  & 3.779 & 0.07 & 0.29 & 16.28 & 31 & 140 & 0.04 \\
26 & SDSS J141754.96$+$270434.4  & 2.723 &\nodata & \nodata & \nodata & \nodata & \nodata & \nodata \\
27 & 2MASX J10304263$+$0418219  & 2.774 & 0.19 & 0.47 &  8.67 & 48 &  25 & 0.14 \\
28 & 2MASX J12374668$+$4812273  & 4.164 & 0.12 & 0.66 & 10.39 & 31 & 192 & 0.11 \\
29 & 2MASX J11410001$+$3848078  & 4.108 & 0.12 & 0.55 & 10.62 & 37 &   3 & 0.04 \\
30 & SDSS J153618.97$+$452246.8  & 3.749 & 0.11 & 0.46 & 16.36 & 44 &  81 & 0.08 \\
31 & 2MASX J09362208$+$3906291  & 4.276 & 0.07 & 0.50 & 11.43 & 30 &  -4 & 0.05 \\
32 & SDSS J140138.37$+$263527.6   & 4.287 & 0.09 & 0.13 & 14.41 & 40 &  11 & 0.04 \\
33 & SDSS J154950.91$+$234444.1   & 4.049 & 0.18 & 0.19 & 12.20 & 36 & 168 & 0.06 \\
34 & 2MASX J12321515$+$1021195   & 2.841 & 0.16 & 0.02 & 13.18 & 42 &  10 & 0.10 \\
35 & 2MASX J12005393$+$4800076  & 4.226 & 0.18 & 0.50 &  8.51 & 36 & 145 & 0.05 \\
36 & 2MASX J07380615$+$2823592  & 3.687 & 0.13 & 0.66 & 12.61 & 45 &  31 & 0.06 \\
37 & 2MASX J16014061$+$2718161  & 2.823 & 0.04 & 0.70 & 11.43 & 59 & 180 & 0.23 \\
38 & SDSS J143447.86$+$020228.6   & 4.244 & 0.24 & 0.10 & 11.61 & 63 &  10 & 0.09 \\
39 & SDSS J112928.74$+$025549.9   & 3.788 & 0.09 & 0.39 & 14.45 & 13 & 221 & 0.05 \\
40 & SDSS J101603.97$+$303747.9   & 3.934 & 0.17 & 0.68 & 10.03 & 12 & 107 & 0.04 \\
41 & 2MASX J00155012$-$1002427   & 2.982 & 0.36 & 0.36 &  9.81 & 14 &  38 & 0.08 \\
42 & 2MASX J13103930$+$2235023  & 3.690 & 0.13 & 0.69 &  9.73 & 36 &  30 & 0.06 \\
43 & 2MASX J13475962$+$3227100  & 3.590 & 0.14 & 0.16 & 14.17 & 37 &  97 & 0.05 \\
44 & 2MASX J10041606$+$2958441  & 4.439 & 0.17 & 0.30 &  6.74 & 48 & 137 & 0.09 \\
45 & 2MASX J10095635$+$2611324  & 3.804 & 0.08 & 0.70 & 11.82 &  1 & 105 & 0.11 \\
46 & 2MASX J09334777$+$2114362  & 2.929 & 0.22 & 0.38 &  5.48 &  6 & 140 &0.06 \\
47 & 2MASX J14472834$+$5908314  & 3.857 & 0.14 & 0.69 & 11.70 & 21 &  48 & 0.08 \\
48 & 2MASX J13435549$+$2440484  & 2.427 & 0.55 & 0.00 & 11.14 & 33 & 159 & 0.05 \\
49 & 2MASX J13422833$+$1157345  & 4.234 & 0.40 & 0.18 &  8.76 &  2 &  53 & 0.06 \\
50 & CGCG 122-067                            & 1.663 & 0.34 & 0.41 &  9.01 & 38 &  84 & \nodata \\
51 & SDSS J121644.34$+$122450.5  & 3.992 & 0.06 & 0.70 & 12.23 & 21 & 125 & 0.10 \\
52 & SDSS J040422.91$-$054134.9   & 3.917 & 0.31 & 0.30 & 14.75 & 39 & 100 & 0.03 \\
53 & 2MASX J00380781$-$0109365   & 3.409 & 0.39 & 0.43 & 13.60 & 58 &  76 & 0.07 \\
\enddata
\tablenotetext{a}{Scale [kpc/$''$].}
\tablenotetext{b}{Bulge fraction in SDSS DR7 $r$-band image.}
\tablenotetext{c}{Bulge eccentricity.}
\tablenotetext{d}{Disk exponential scale length.}
\tablenotetext{e}{Disk inclination ($\arcdeg$).}
\tablenotetext{f}{Disk PA ($\arcdeg$).}
\tablenotetext{g}{Smoothness in r band.}

\end{deluxetable}
\clearpage

\begin{deluxetable}{clcccllcll}
\tablecaption{Candidate Cluster and Group Membership}
\tablehead{
\colhead{SS}  &\colhead{NED Name} & \colhead{Redshift} & \colhead{N1\tablenotemark{a}} & \colhead{N10\tablenotemark{b}} & \colhead{Cluster Name} & \colhead{Type}& \colhead{Redshift} & \colhead{ztype\tablenotemark{c}} 
                       &  \colhead{Sep($'$)}}
\startdata
02 & 2MASX J10405643-0103584   &  0.250303    & 1  &  8  & SDSS CE J160.241898-01.069106 & GClstr & 0.254019 & EST & 0.013 \\
04 & 2MASX J10100707+3253295  &  0.289913    & 2  & 17 & GMBCG J152.52936+32.89139      & GClstr & 0.319000 & PHOT & 0.001 \\
09 & 2MASX J09094480+2226078  &  0.285386    & 1  &   9 & GMBCG J137.43670+22.43538      & GClstr & 0.303000 & PHOT & 0.000 \\
12 & 2MASX J09260805+2405242  &  0.222451    & 1  & 22 & WHL J092608.1+240524                 & GClstr & 0.178000 & PHOT & 0.000 \\
13 & 2MASX J17340613+6029190  &  0.275807    & 1  &  2  & SDSSCGB 59704                            & GGroup & 0.276000\tablenotemark{d} & SPEC & 0.450 \\ 
17 & 2MASX J11535621+4923562  &  0.166892    & 3  &  69 & OGC  0586 CLUSTER                    & GClstr   & 0.166187  & SPEC & 0.000 \\ 
35 & 2MASX J12005393+4800076  &  0.278617    & 1  & 13 & GMBCG J180.22479+48.00211      & GClstr & 0.252000 & PHOT & 0.001 \\
37 & 2MASX J16014061+2718161  &  0.164554    & 3  &163 & GMBCG J240.41924+27.30444      & GClstr & 0.193000 & PHOT & 0.000 \\
43 & 2MASX J13475962+3227100  &  0.223113    & 1   & 13 &SDSSCGB 16827                            & GGroup & \nodata \tablenotemark{d} & \nodata & 0.748 \\ 
50 & CGCG 122-067                        &  0.089008    & 5   &302 &MSPM 05544                                   & GClstr & 0.089190 & SPEC & 0.001 \\
\enddata
\tablenotetext{a}{Number of galaxies within 1 Mpc and 500 km s$^{-1}$.}
\tablenotetext{b}{Number of galaxies within 10 Mpc and 5000 km s$^{-1}$.}
\tablenotetext{c}{Redshift type, from NED. EST--estimated, PHOT--photometric, and SPEC--spectroscopic.}
\tablenotetext{d}{The association of the super spiral galaxy with the compact group is based only on proximity on the sky. 
                            The group redshift in NED for SDSSCGB 59704 appears to be based only on the redshift of the super spiral.
                             None of the other galaxies in SDSSCGB 16827 have measured redshifts.}
\end{deluxetable}

\appendix{}

\section{Custom Photometry and Validation of $M_\mathrm{stars}$ and SFR}

\begin{figure}[ht]
   \plotone{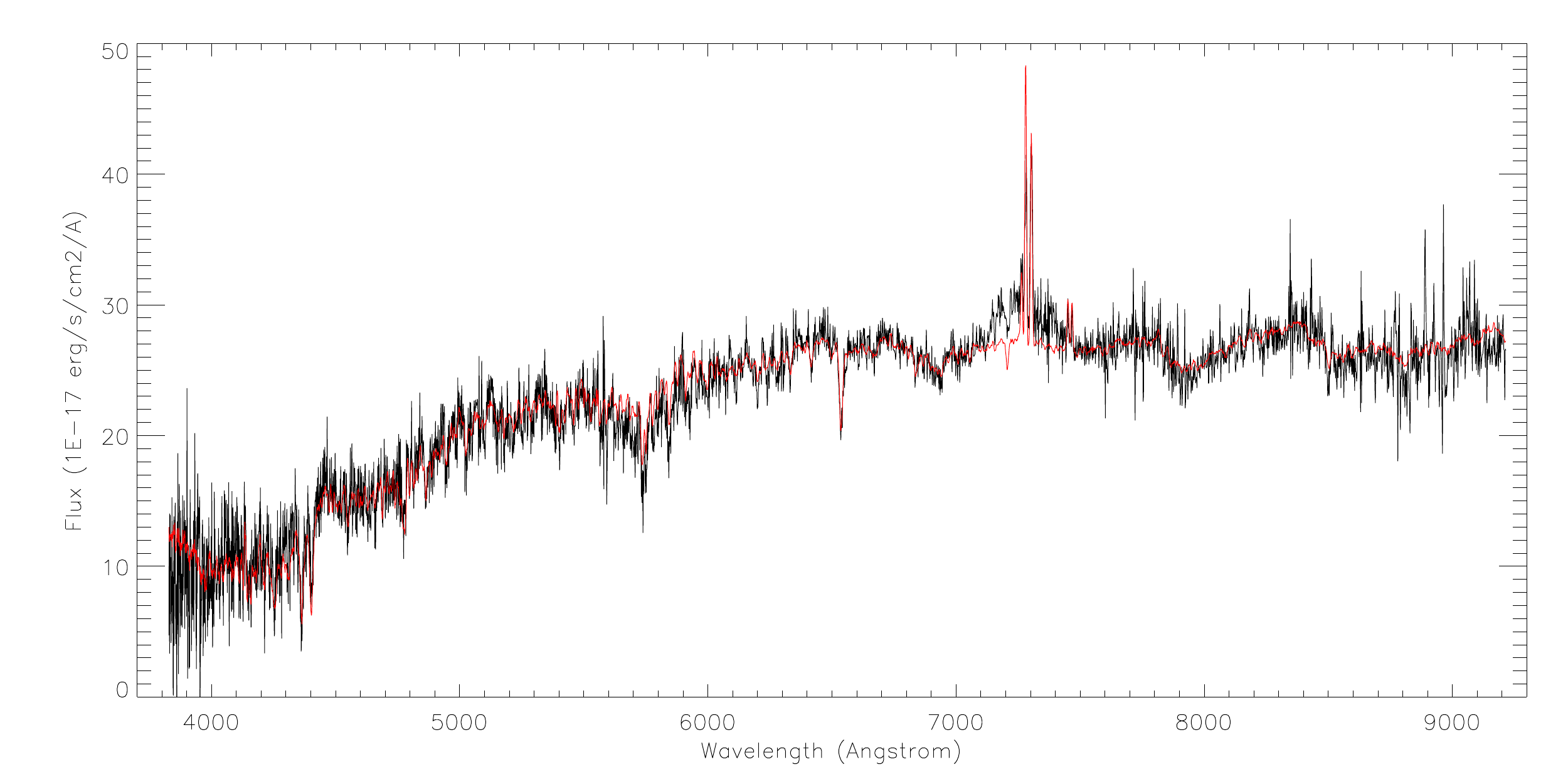}
   \figcaption{SDSS DR9 optical spectrum and spectral model of SDSS J094700.08+254045.7 (SS 16). Note the possible broad H$\alpha$ emission
   line not fit by the spectral model, indicative of a Seyfert 1 AGN
   \label{9}.}
\end{figure}

\begin{figure}[ht]
   \plotone{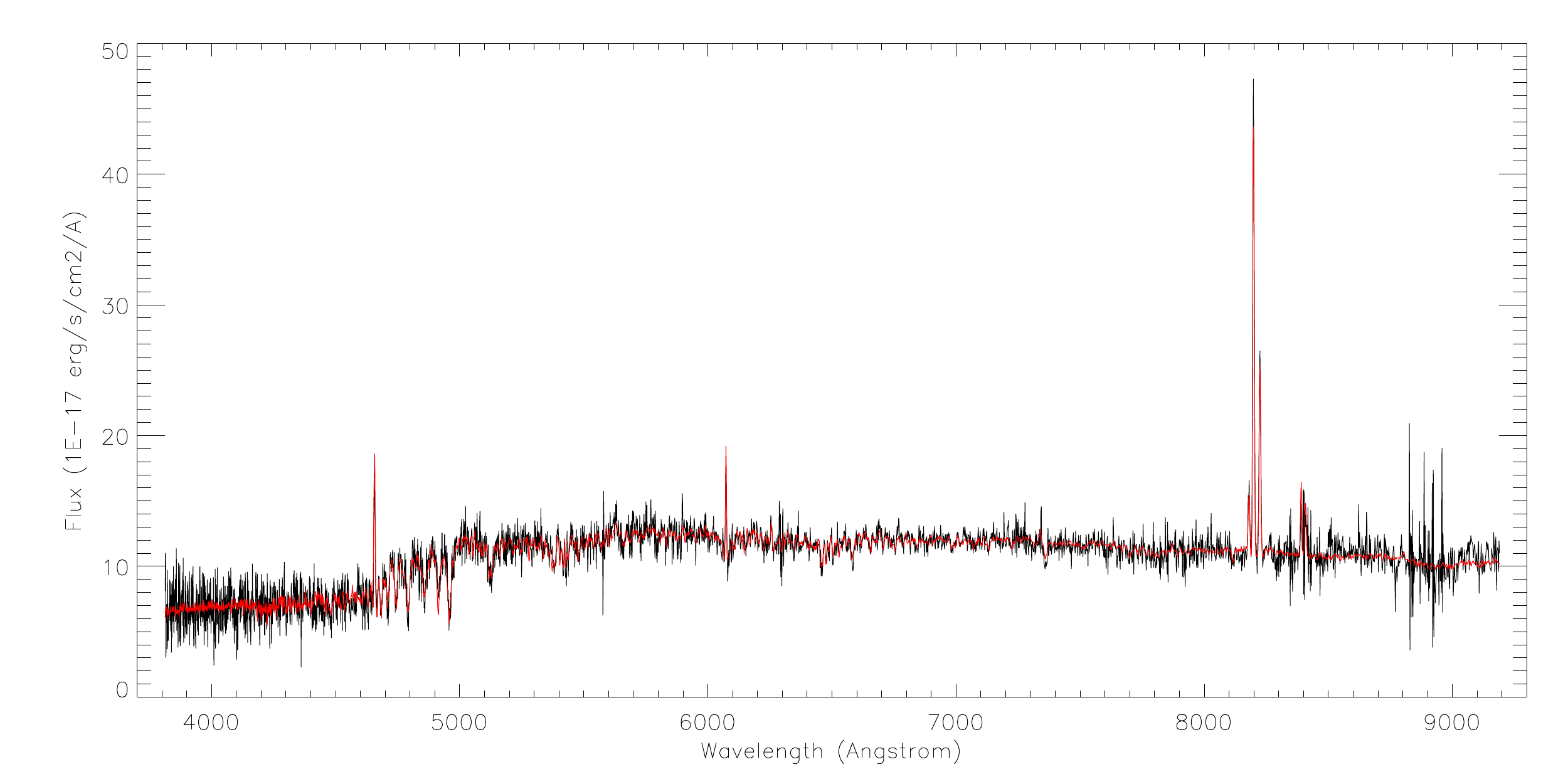}
   \figcaption{SDSS DR9 optical spectrum and spectral model of 2MASX J13275756+334529 (SS 05). The blue spectral slope and strong H$\alpha$ emission
                     indicate starburst activity in the galaxy bulge that contributes to the high star formation rate in the galaxy as a whole.
                     \label{10}}
\end{figure}

In order to validate our stellar mass and SFR estimates, which are based on $Ks$, $u$, $r$,  and {\it WISE} 12 $\mu$m photometry, we make a more
detailed analysis of two representative examples from our super spiral sample. We remeasure their photometry in matched apertures, rather
than relying on catalog photometry. Then we fit their SEDs to make full use of the available multi-band photometry 
to estimate more accurate $M_\mathrm{stars}$ and SFR. We chose SDSS J094700.08+254045.7  (SS 16) for this analysis because it is one of the brightest super spirals in 
our sample, with good photometry in many bands, and typical colors. The SDSS spectrum of its bulge is also typical of
most super spirals, being dominated by an old stellar population (Figure \ref{9}).  We also make a  detailed study of 2MASX J13275756+3345291 (SS 05), which is the most 
luminous (non-QSO) mid-IR source in our sample and has an SDSS nuclear spectrum with strong young stellar component and high-equivalent width H$\alpha$ emission (Figure \ref{10}), characteristic 
of starburst activity.

\begin{figure}[ht]
   \plotone{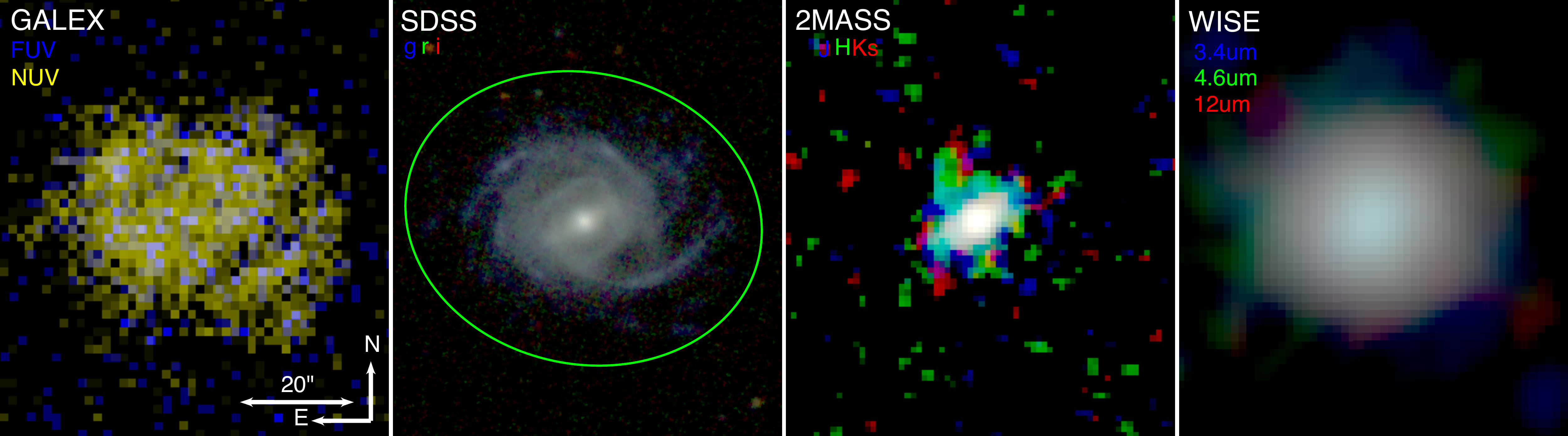}
   \figcaption{{\it GALEX}, SDSS, 2MASS and {\it WISE} images of SDSS J094700.08+254045.7 (SS 16). The image scale is 1.991 kpc/$''$. 
   The photometric aperture with major and minor axes of
   125 kpc  and 102 kpc, respectively, is indicated by the ellipse on the SDSS image.
    \label{11}}
\end{figure}

We remeasured {\it GALEX} (FUV, NUV), SDSS ($u$, $g$, $r$, $i$, $z$), 2MASS ($J$, $H$, $Ks$) and {\it WISE} band 1-4 photometry for SS 16 using the SAOImager 
ds9 \citep{jm03} on images retrieved from MAST, SDSS, and IRSA (Figure \ref{11}).  Aperture and color corrections were applied as necessary and the {\it GALEX} 
and SDSS photometry was corrected for foreground extinction due to the Milky Way dust \citep{wtm05,slb02}. The Galactic extinction is
a modest $A_V=0.063$ mag (NED). We used an elliptical aperture with semimajor and semiminor axes of  $31\farcs5$  and $25\farcs5$, respectively, in order to capture the 
full flux of the spiral disk in all bands. This corresponds to major and minor diameters of 125 kpc and 102 kpc.   We also compute $3\sigma$ {\it IRAS} upper limits based on the rms
uncertainty measured by IRSA's Scan Processing and Integration tool (SCANPI) to constrain the FIR luminosity. 

\begin{figure}[ht]
   \plotone{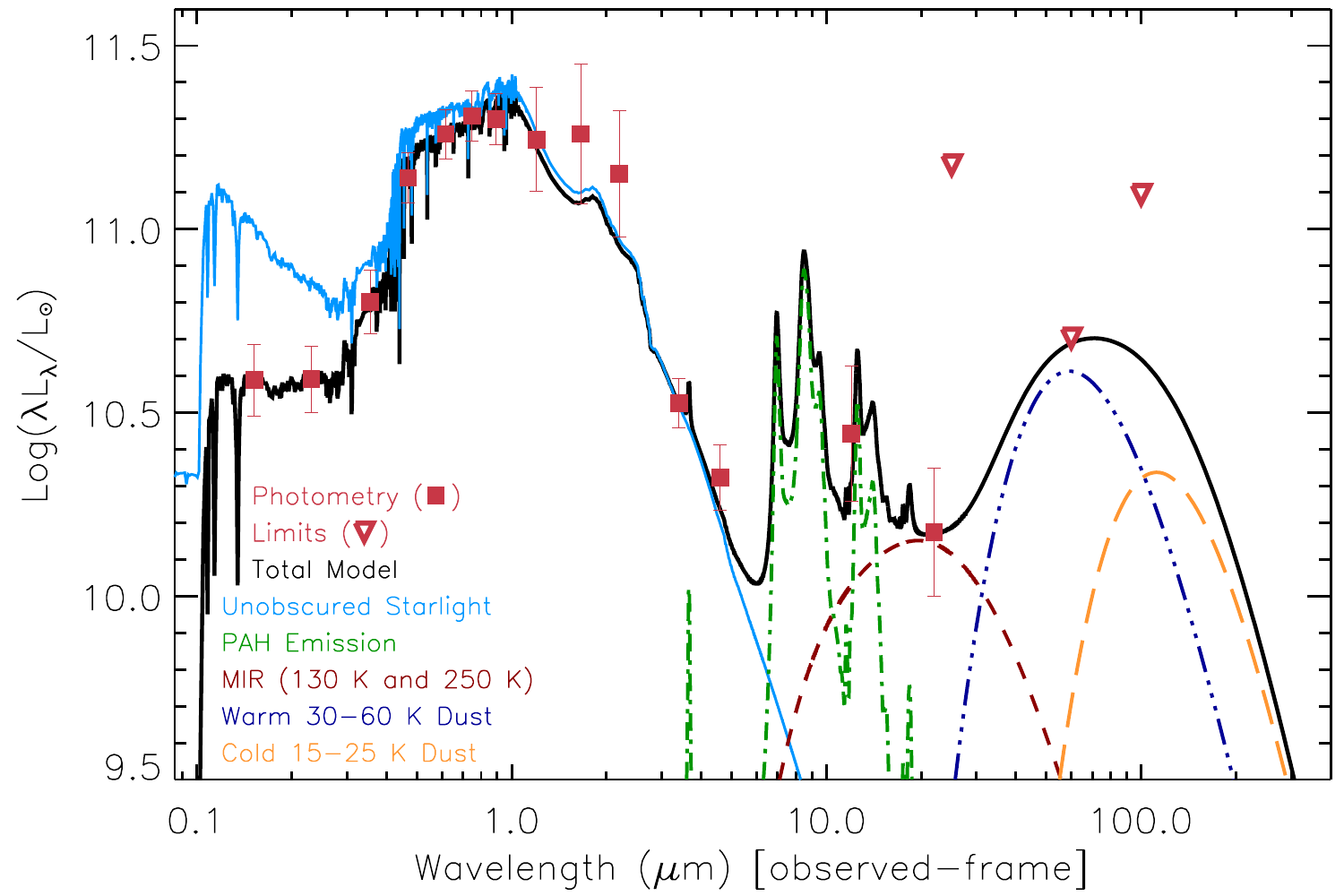}
   \figcaption{Spectral Energy Distribution of SDSS J094700.08+254045.7 (SS 16) fit by  {\sc magphys}. {\it GALEX} (NUV, FUV), SDSS ($u$, $g$, $r$, $i$, $z$), 2MASS ($J$, $H$, $Ks$), and {\it WISE}
                     band 1-4 photometry are measured in the aperture shown in \ref{11}. {\it IRAS} upper limits at 25, 60, and 100 $\mu$m are estimated using SCANPI.
                     \label{12}}
\end{figure}

We present the SED of SS 16 in Figure \ref{12}. The galaxy is detected in all {\it GALEX}, SDSS, 2MASS, and {\it WISE} bands,
but is undetected by {\it IRAS}. The  UV through near-IR data points reveal a massive old stellar population plus a young stellar population.
Mid-IR emission appears to be dominated by PAHs and warm dust from star formation. We fit the SED using {\sc magphys} template fitting \citep{dce08}. This gives a total stellar 
mass of  $1.8^{+0.3}_{-0.2}\times10^{11} M_\odot$ and SFR of $9.9^{+1.6}_{-0.3} M_\odot$ yr$^{-1}$. We get a consistent estimate of $1.2 \pm 0.1 \times 10^{11} M_\odot$ 
for the stellar mass from the $u-r$ color and $Ks$-band luminosity, applying the \cite{bmk03} prescription for color-dependent mass-to-light ratio (Table 1). The {\it WISE} band 3 luminosity gives a consistent 
SFR of $13.5 \pm 0.2  M_\odot$ yr$^{-1}$, using the prescription of \cite{cvc15}. 
Lacking FIR detections, we do not have a good handle on the total dust mass, however, the SED fit formally yields a dust mass of $\sim10^8 M_\odot$, based on the 
PAH emission and FIR upper limits. This corresponds to roughly $\sim10^{10} M_\odot$ of gas, assuming a standard gas/dust ratio of 100.

\begin{figure}[ht]
   \plotone{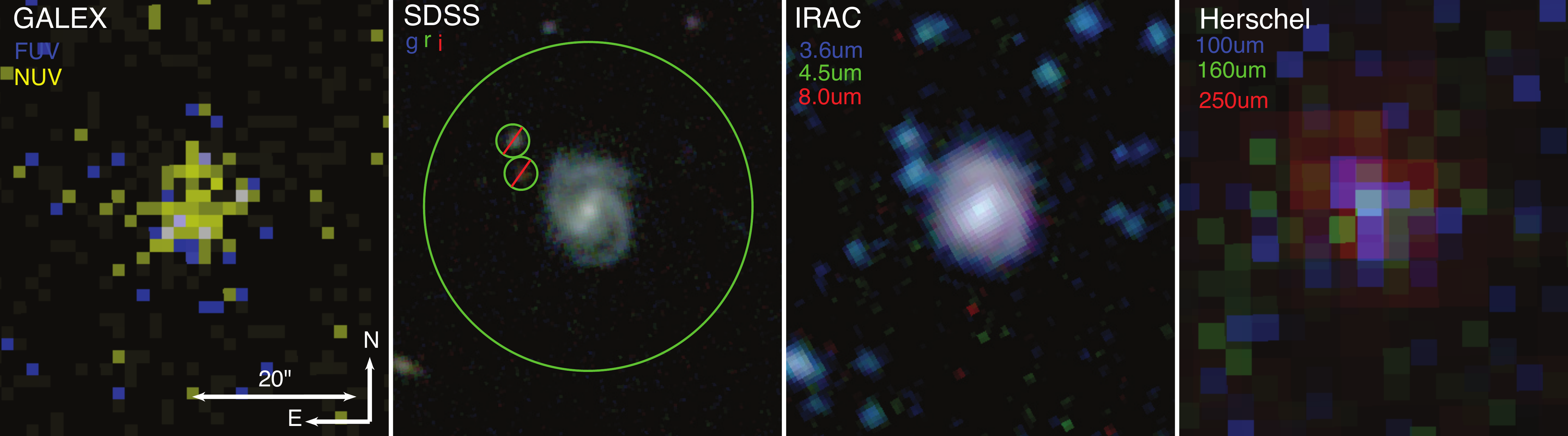}
   \figcaption{{\it GALEX}, SDSS, {\it Spitzer} IRAC, and {\it Herschel} images of 2MASX J13275756+334529 (SS 05). The image scale is 3.898 kpc/$''$. 
   The circular photometric aperture with diameter 156 kpc and two exclusion regions are indicated on the SDSS image.
    \label{13}}
\end{figure}

We remeasured {\it GALEX} (FUV, NUV), SDSS ($u$, $g$, $r$, $i$, $z$), 2MASS ($J$, $H$, $Ks$) and {\it WISE} band 1-4 photometry for  SS 05 (Figure \ref{13}), using
a similar procedure.  We also retrieved {\it Spitzer} IRAC and MIPS, and {\it Herschel} PACS and SPIRE images from the respective IRSA and ESA archives to measure the IR fluxes. 
We used a circular aperture with $20\farcs0$  (156 kpc) radius for most bands. However, at SPIRE wavelengths, we used the larger point source apertures of $22''$, $30''$, and $42''$, 
in order to contain the broader point-spread function. The Galactic extinction is only $A_V=0.034$ mag (NED). 

\begin{figure}[ht]
   \plotone{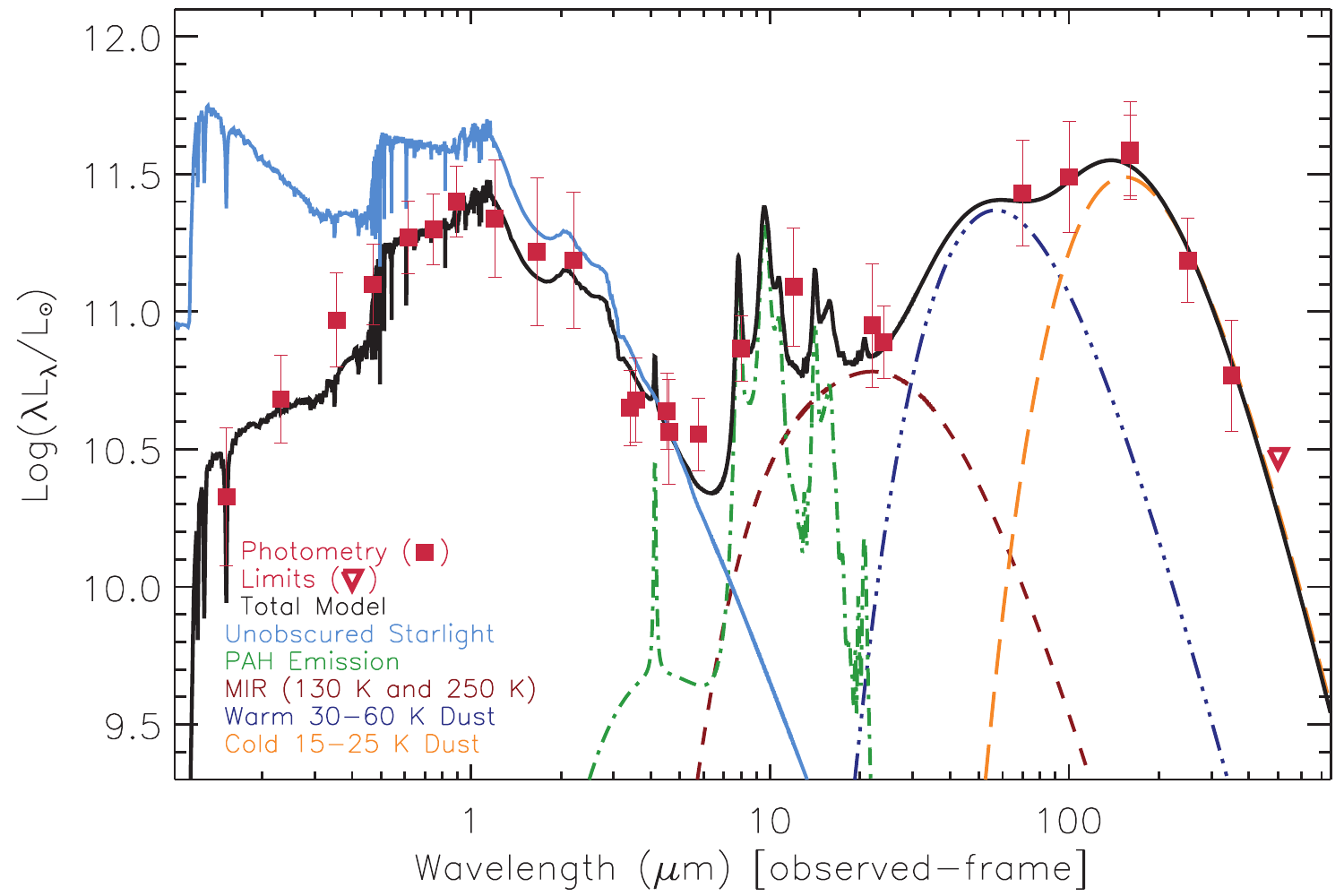}
   \figcaption{Spectral Energy Distribution of 2MASX J13275756+334529 (SS 05) fit by  {\sc magphys}. {\it GALEX} (NUV, FUV), SDSS ($u$, $g$, $r$, $i$,  $z$), 2MASS ($J$, $H$, $Ks$), 
                     {\it Spitzer} IRAC and MIPS 24, 70, and 160 $\mu$m, {\it WISE} band 1-4, {\it Herschel} PACS 100 and 160 $\mu$m, and SPIRE photometry are measured in the aperture shown 
                      in Figure \ref{13}. SPIRE 500 $\mu$m luminosity is an upper limit.
                      \label{14}}
\end{figure}

We present the SED of SS 05 in Figure \ref{14}. The galaxy is detected in all measured bands except the SPIRE 500 $\mu$m band.
In contrast to SS 16, there is a stronger component of emission from young stars, and much more luminous 
IR emission from star formation activity. We fit the SED using {\sc magphys}, yielding a total stellar mass of  $2.04^{+0.05}_{-0.09} \times10^{11} M_\odot$ and SFR
of $40.5^{+6.5}_{-0.5} M_\odot$ yr$^{-1}$. The stellar mass  is consistent with the value of $1.6 \pm 0.3 \times 10^{11} M_\odot$ that we obtain from the $u-r$ color and $Ks$-band 
luminosity (Table 1). The {\it WISE} [12] luminosity gives a somewhat larger SFR of $65 \pm 4 M_\odot$ yr$^{-1}$, using the conversion factor of \cite{cvc15}.  
The {\it Herschel} FIR measurements yield a secure estimate of total dust mass from the SED fit of $7^{+3}_{-1} \times 10^8 M_\odot$, corresponding to 
$7 \times 10^{10} M_\odot$ of gas, assuming a standard gas/dust ratio of 100.

\clearpage

\clearpage

\end{document}